\documentclass[referee,traditabstract]{aa}
\usepackage{array}
\usepackage{tabularx}
\usepackage{savesym}
\usepackage{amsmath}
\savesymbol{iint}
\usepackage{txfonts}
\restoresymbol{TXF}{iint}
\usepackage[T1]{fontenc}
\usepackage{longtable}
\usepackage{natbib}
\usepackage{epsfig}
\usepackage[dvips]{color}
\usepackage{graphics}
\usepackage{float}
\usepackage{graphicx}

\def\apjl{Astrophys. J. Letters}
\def\apjs{Astrophys. J. Suppl.}

\begin{document}

\title{Potential biosignatures in super-Earth atmospheres}

\subtitle{I. Spectral appearance of super-Earths around M dwarfs}

\author{H. Rauer \inst{1,2}
          \and
S. Gebauer \inst{2}
\and
P. v. Paris\inst{1}, J. Cabrera \inst{1}
  \and
M. Godolt\inst{2},
  J.L. Grenfell \inst{2}
\and
A. Belu\inst{3,4},
  F. Selsis\inst{3, 4}
\and
 P. Hedelt \inst{3,4} 
\and F. Schreier \inst{5}}

\institute{Institut f\"ur Planetenforschung, Deutsches
  Zentrum f\"ur Luft- und Raumfahrt, Rutherfordstra{\ss}e 2, 12489
  Berlin, Germany, \email{heike.rauer@dlr.de}
\and
Zentrum f\"ur Astronomie und Astrophysik,
  Technische Universit\"at Berlin, Hardenbergstra{\ss}e 36, 10623
  Berlin, Germany
\and
Universit\'e de Bordeaux, 
Observatoire Aquitain des Sciences de l'Univers, 
2 rue de l'Observatoire, 
BP 89, 
F-33271 Floirac Cedex, France
\and
CNRS, UMR 5804, Laboratoire d'Astrophysique de Bordeaux, 2 rue de l'Observatoire, BP 89, F-33271 Floirac Cedex, France
\and
Institut f\"ur Methodik der Fernerkundung, Deutsches
  Zentrum f\"ur Luft- und Raumfahrt 82234 Oberpfaffenhofen-Wessling, Germany}

\date{}
\abstract{Atmospheric temperature and mixing ratio profiles of
  terrestrial planets vary with the spectral energy flux distribution
  for different types of M-dwarf stars and the planetary gravity.  We investigate the resulting
  effects on the spectral appearance of molecular absorption bands, which are
  relevant as indicators for potential planetary habitability during
  primary and secondary eclipse for transiting terrestrial planets
  with Earth-like biomass emissions.  Atmospheric profiles are
  computed using a plane-parallel, 1D climate model coupled with a
  chemistry model. We then calculate simulated spectra using a
  line-by-line radiative transfer model.

 We find that emission spectra during secondary eclipse show
    increasing absorption of methane, water, and ozone for planets
    orbiting quiet M0-M3 dwarfs and the active M-type star AD Leo compared with solar-type central stars. However,
    for planets orbiting very cool and quiet M dwarfs (M4 to M7),
    increasing temperatures in the mid-atmosphere lead to reduced
    absorption signals, which impedes the detection of molecules
    in these scenarios. Transmission spectra during primary
    eclipse show strong absorption features of CH$_4$, N$_2$O and
    H$_2$O for planets orbiting quiet M0-M7 stars and AD Leo. The
    N$_2$O absorption of an Earth-sized planet orbiting a quiet M7 star can even be
    as strong as the CO$_2$ signal.  However, ozone absorption
    decreases for planets orbiting these cool central stars owing to
    chemical effects in the atmosphere. To investigate the effect on
  the spectroscopic detection of absorption bands with potential
  future satellite missions, we compute signal-to-noise-ratios (SNR)
  for a James Webb Space Telescope (JWST)-like aperture telescope.}

\keywords{   Planets and satellites: atmospheres; Astrobiology}

\maketitle

\section{Introduction}

The majority of the more than 500 extra-solar planets
found so far are giant gas planets.
However, with improving detection techniques, more and more low-mass planets will be found. At the time
of writing this paper, we know of more than 20
planets with (minimum) masses below 10 $M_{\oplus}$. Two systems
(around Gliese 581 and HD 40307) host three potential low-mass planets
each (\citealp{Udry2007}, \citealp{mayor2009gliese},
\citealp{mayor2009triple}). The only two low-mass planets where both
radius and mass could be measured are CoRoT-7b (\citealp{leger2009},
\citealp{queloz2009}), a transiting planet of approximately 5
$M_{\oplus}$ and 1.7 $R_{\oplus}$, and GJ 1214b \citep{charb2009},
with 6.5 $M_{\oplus}$ and 2.7 $R_{\oplus}$. CoRoT-7b, which is likely
to be a rocky planet, may therefore be classified as terrestrial (a so
called super-Earth).

Whereas atmospheres of close-in gas giant planets around bright stars
have been already spectroscopically characterised with the Hubble
Space Telescope (HST) and the Spitzer IR-telescope (e.g.
\citealp{vidal2004}, \citealp{tinetti2007}, \citealp{swain2008},
\citealp{knutson2008}, \citealp{swain2009},
\citealp{knutson2009tres_tinvers}), terrestrial extra-solar planets
seem out of reach with currently available telescopes. However, future
satellite missions, such as the JWST (James Webb Space Telescope) or proposed
dedicated missions for spectral characterisation, will provide an
improved detection range. To which extent their performance will be sufficient to  spectroscopically characterise the atmosphere of terrestrial planets in the habitable zone of their central star is the
subject of increasing modelling investigations.

Previous major modelling efforts producing synthetic spectra of
hypothetical terrestrial extra-solar planets were performed by
e.g. \citet{selsis2000}, \citet{selsis2002}, \citet{DesMarais2002},
\citet{Segura2003}, \citet{Segura2005}, \citet{Tinetti2006},
\citet{ehrenreich2006}, \citet{Kaltenegger2007},
\citet{Kaltenegger2009} and \citet{kaltenegger2010}. They examined the
influence of e.g. stellar type, atmospheric abundances, background
atmospheres or atmospheric evolution on the emission and transmission
spectra.

This work focuses on emission and transmission spectra of
  super-Earth planetary atmospheres that orbit in the habitable zone
  (HZ) around M-dwarf stars because in these cases the highest
  signal-to-noise-ratios (SNRs) for potentially habitable planets are
  expected. Spectral resolution and the expected SNRs are studied for
  a photon-limited telescope with 6.5 m aperture (JWST-sized). Real
  SNRs will be significantly lower because of instrumental noise
  and the astronomical background. The values here serve as a
  "best case" therefore and provide a first estimate on the possible detection
  performance. A parametric study of expected SNRs for a wide range of
  star-planet combinations including instrumental noise for the JWST
  has been performed by \citet{Belu2010}. We complement this study by
  a detailed investigation that includes the effects of atmospheric
  chemistry on habitable super-Earth planets.

Our approach is similar to that of \citet{selsis2000}, \citet{Segura2003} and
\citet{Segura2005}. We first apply an atmosphere column model to
calculate the mean temperature profile and the
corresponding chemical profiles assuming an Earth-like planet
development and biomass emissions.  We chose this approach because the
T-p-profiles as well as the atmospheric mixing ratios vary depending on
stellar type and planetary parameters. This was first shown for
  planets orbiting M-dwarf stars by \citet{Segura2005}, who
  demonstrated the effect of the changing atmospheric chemistry on
  emission flux spectra of these Earth-like extra-solar planets. This early work
  did not include transit transmission spectra and SNR
  estimates or a change of planetary gravity though. SNRs were later calculated for a 6.5m JWST-aperture
  telescope by \citet{Kaltenegger2009} in a simplified approach, which
  kept the planetary atmospheric conditions fixed to those of a modern Earth when
  changing the central star. Their results showed the
  difficulty to detect the weak biosignatures in transmission spectra
  with the JWST, but also discussed how this could be overcome by
  co-adding several transit measurements. Our main goal is to extend these SNR investigations 
  by including the effect of chemistry to study whether the
  detection of biosignatures could be favoured by chemical processes, as suggested by \citet{Segura2005}.
In addition, we also consider transmission and emission spectra.

The paper is organised as follows: Section 2 describes the
  atmospheric and radiative transfer models and the modelled
  scenarios. In Sect. 3 we show how the SNR calculations are
  performed.  Section 4 compares the resulting emission and transmission
spectra with modern Earth spectra, followed by the
resolution and SNR calculations. Section 5 summarises the
results.


\section{Models and simulated scenarios}

\subsection{Atmospheric and spectral models}
\label{1dmodel}

We used a 1D radiative-convective photochemical model to
calculate the chemistry and climate of planets with an Earth-like atmospheric chemical composition. The model
calculates the globally, diurnally-averaged atmospheric temperature,
pressure, and composition profiles for cloud-free conditions.  The
original code has been described in detail by \citet{Kasting84} and
was further developed by \citet{Segura2003}, \citet{Grenfell2007asbio}
and \citet{Grenfell2007}. Here, we briefly summarise the
  model and describe its recent improvements.

The model extends from the ground up to the
  mid-mesosphere. The climate module grid is divided into 52 layers
    on a log-pressure grid.  Compared with
  \citet{Grenfell2007asbio}, we improved the calculation of the grid. Depending on how much water
    is evaporating at the surface, the pressure is re-adjusted in each
    iteration step. The chemistry module is divided into 64
    equidistant layers in altitude. As an upgrade to previous model versions, the top 
layer is now equal to the top-of-atmosphere (TOA) altitude derived by the climate
    module. Hence, the model grid consistently takes into account the
  changing scale height with varying gravity  in the
  climate and chemical modules, allowing the simulation of super-Earth
  planets with high gravity with consistent resolution.

The radiative transfer for incoming stellar radiation in our
  climate module covers the wavelength region from 237.6 nm to 4.545 $\mu$m with 38 wavelength
  bands. We used a two-stream radiative transfer method based on
  \citet{Toon1989}. The radiatively active species considered are 
   H$_2$O, CO$_2$, O$_2$, O$_3$, and CH$_4$. Absorption
  coefficients are calculated by an exponential sum correlated-k
    method using up to four terms per sum. Rayleigh scattering is
  included for N$_2$, O$_2$, and CO$_2$. 

For thermal emission we considered the wavelength range from
  3.33 -- 1000 $\mu$m in 16 bands, using the Rapid Radiative Transfer Model (RRTM, \citealp{Mlawer1997}). The model includes H$_2$O,
  CO$_2$, O$_3$, CH$_4$, and N$_2$O as radiative gases in the
  IR-range. The absorption coefficients were calculated with the 16-term
  correlated-k method. Because k-coefficients are always calculated for a certain
  composition, temperature, and pressure range, the model is valid
  strictly only within 0.01 - 1050 mbar and for the Earth's atmospheric
    temperature profile $\pm$30 K. For this reason, we
  limit our coolest input stellar spectra up to M7 dwarf stars for Earth-sized planets, and up to M5-type stars for super-Earth planets.  Cooler M dwarfs would result in T-p profiles widely
  outside the applicable model range. Wet or dry adiabatic H$_2$O
  lapse rates are used to obtain the temperature profile in the
  troposphere where applicable.  The water profile is taken from
  \citet{MW1967}. We consider a mean solar zenith angle of
  60$^{\circ}$ for the climate model.

The chemistry module includes 55 species with 212
  chemical reactions. As in most previous model approaches and for
  modern Earth simulations, CO$_2$, O$_2$ and N$_2$ are
  isoprofiles. To calculate the photolysis rates, we took
  into account a wavelength range from 121.4 - 855 nm and considered
  O$_3$, O$_2$, CO$_2$, H$_2$O, SO$_2$, CH$_4$, and H$_2$S as
  radiative gases. Rayleigh scattering was included for N$_2$, O$_2$,
  and CO$_2$. The mean solar zenith angle is 45$^{\circ}$, following
  the approach by \citet{Segura2005}. In addition to the species
  considered by \citet{Segura2005}, we included H$_2$O and CH$_4$ when calculating 
the mean molecular weight of air and the concentration of
  the fill gas nitrogen. We furthermore changed the convergence
  criterion for the numerical solution from previous model versions to
  a maximum difference in volume mixing ratio for all species in all
  model layers of less than 10$^{-4}$ to ensure convergence for all atmospheric profiles. Moreover, the height of the cold
    trap is now given by the climate module, and the mixing ratios of
    all chemical species are adjusted according to how much water is
   evaporated at the surface.

A main objective of this work is to present simulated spectra
  for transiting planets, corresponding to our simulated model
  scenarios. We therefore use the modelled T-p profiles together with
  the corresponding chemical abundances to calculate spectra for
  primary and secondary transit geometry.  The radiative transfer
  model SQuIRRL (Schwarz\-schild Qua\-dra\-ture Infra\-Red Radiation
  Line-by-line, \citealp{schreier2000}) was used to calculate synthetic spectra in the infrared from 1.8 to 20 $\mu$m.
SQuIRRL was found to be comparable with other radiative transfer
codes (e.g.  \citealp{clarmann2003b}, \citealp{melsheimer2005}). The
model takes as input temperature, pressure, and concentration profiles
of the following species from the atmospheric model: H$_2$O, CO$_2$,
CH$_4$, O$_3$, N$_2$O, CO, SO$_2$, NO$_2$, NO, HCl, HNO$_3$,
H$_2$O$_2$, HO$_2$, and CH$_3$Cl. H$_2$O continuum
absorption corrections are performed and local thermodynamic equilibrium
(LTE) is assumed. Absorption cross sections for all molecules except H$_2$O were computed using the Hitran 2004 line parameter data base \citep{HITRAN2004}, for H$_2$O line data from Hitemp 1995 \citep{Rothman95} were used. 
The emission spectra were calculated with a pencil beam looking downwards
with a viewing angle of 38$^{\circ}$ to account for the average
illumination. Transmission spectra are calculated for different
atmospheric heights (62 layers in total) in limb geometry.  The
resulting spectra are binned for resolutions ($R =
\tfrac{\lambda}{\Delta \lambda}$) of $R=5-2000$, corresponding to
instrumental designs for low- and medium-resolution spectroscopy.

\subsection{Modelled scenarios}

\begin{figure}[!h]
\centering  
\includegraphics[width=9.cm]{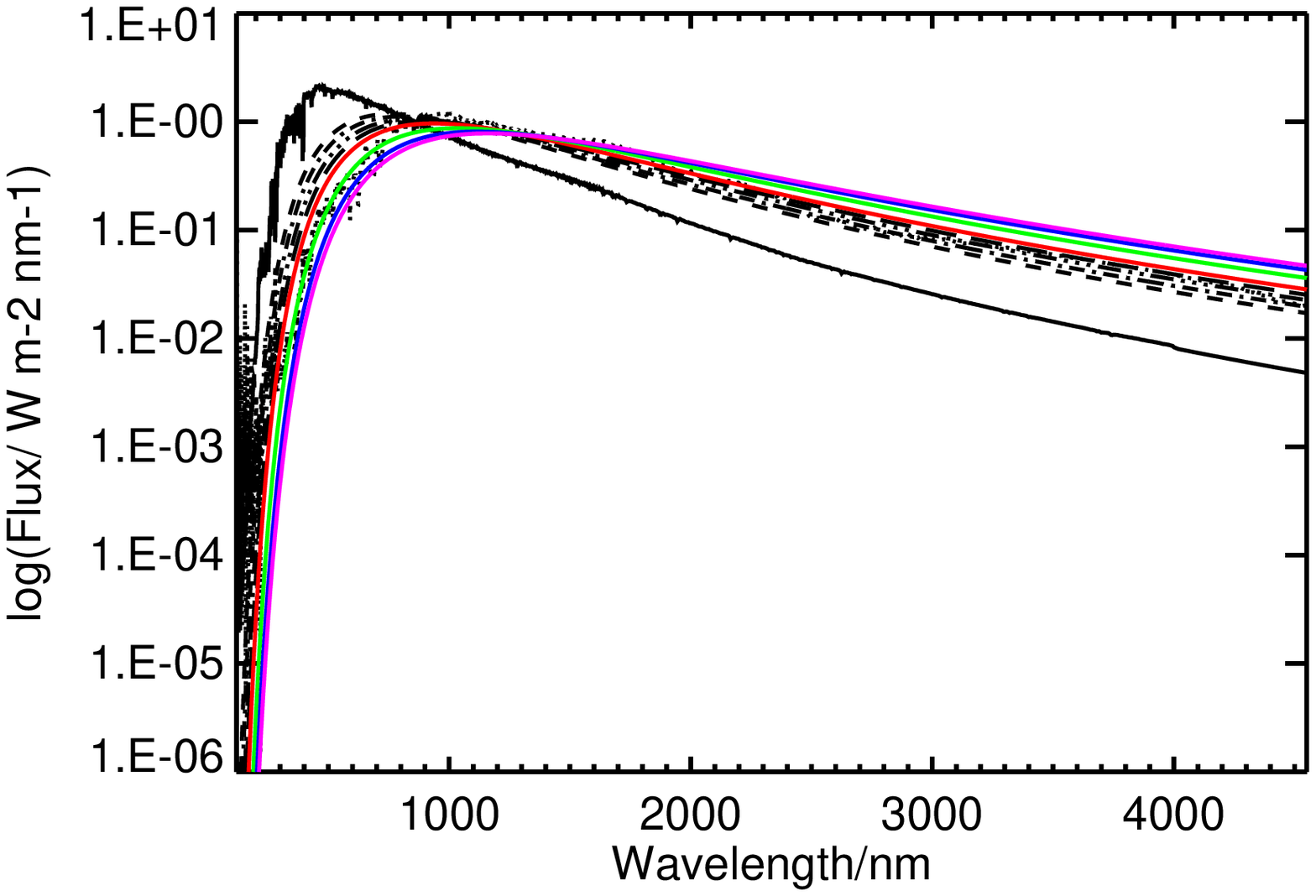} 
\includegraphics[width=9.cm]{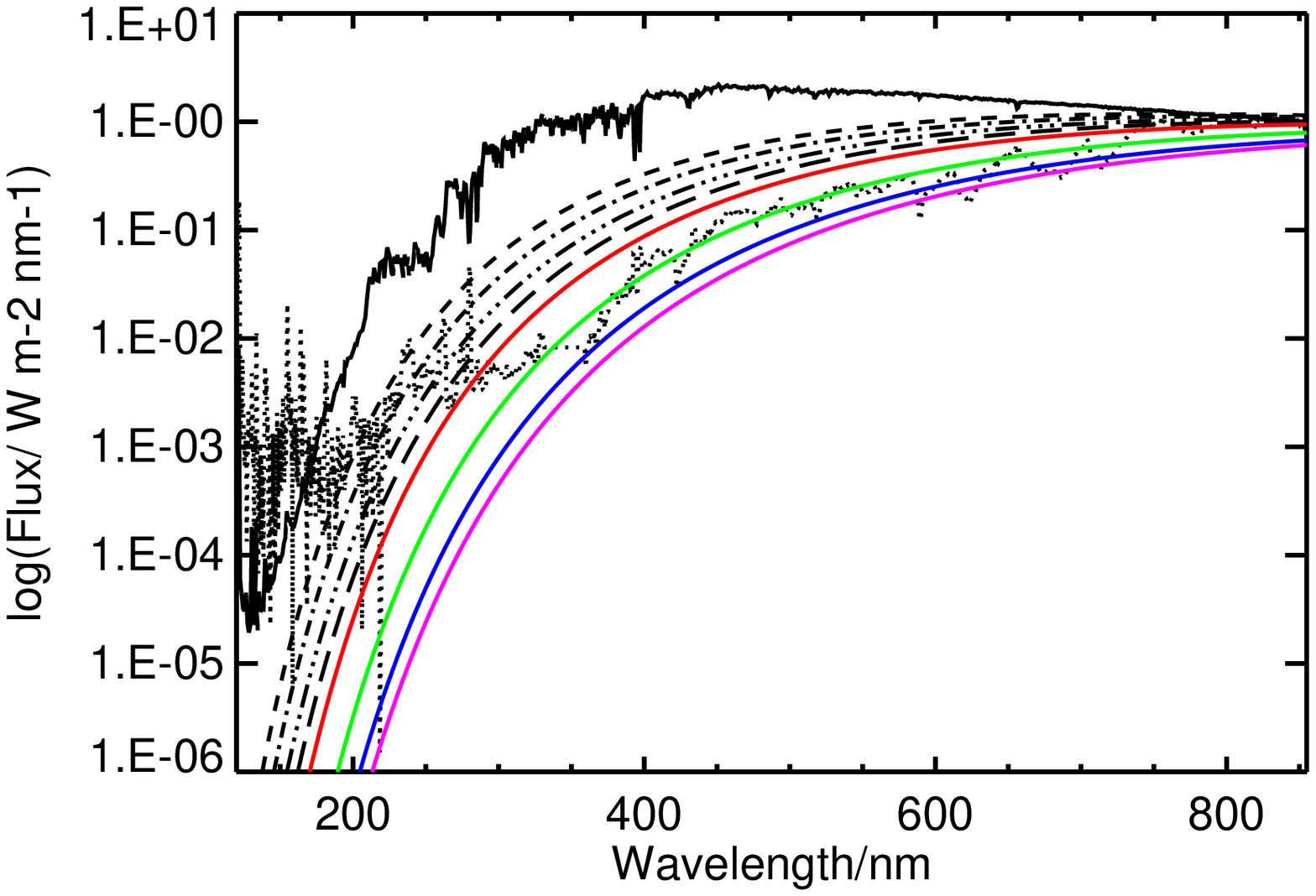}
\caption{Top: Normalised  stellar input
      spectra: Sun (black, solid), AD Leo (dotted), M0 (dashed), M1
      (dash-dot), M2 (dash-dot-dot-dot), M3 (long dashes), M4 (red),
      M5 (green), M6 (blue), and M7 (magenta). Bottom: Zoom into
      the photochemically relevant spectral range.}
\label{inputspectra}
\end{figure}

We study planets with Earth-like chemical composition in the habitable zone of M-dwarf stars of
types M0 -- M7. The spectral appearance of the quiet M-dwarfs is
approximated as black body emission curves according to
effective temperatures and stellar parameters as specified in
\citet{Kaltenegger2009} (see Table \ref{star}). But these input
  spectra neglect that many M-dwarf stars show strong
  activity in the UV range, much more than our Sun, for example. The
  high UV radiation can have a significant effect on atmospheric
  chemistry, as discussed in \citet{Segura2005}. To illustrate the
effect of high stellar UV flux on the expected spectral appearance of
transiting planets, we therefore also used a composite spectrum of AD
Leo (M4.5V) as central star, as described by \citet{Segura2005}. This
high-resolution spectrum is based on observations and a stellar model
atmosphere. Observations were taken from satellite data (IUE,
International Ultraviolet Explorer) and photometry in the visible
\citep{pettersen1989} and near-IR \citep{Leggett1996}. Beyond 2.4
$\mu$m, a stellar atmospheric model (NextGen,
\citealp{hauschildt1999}) was used to calculate a synthetic spectrum.
In addition, the Sun was used as a reference case for comparison. The
solar spectrum was constructed using measurements provided by
\citet{Gueymard2004}. The high-resolution spectra were binned
appropriately for our broadband climate and chemical code.  The
stellar parameters are summarised in Table \ref{star} and the
  input spectra used are shown in Fig. \ref{inputspectra}.  Note the
  significantly higher UV flux of AD Leo compared with the quiet M4 and
  M5 dwarfs shortwards of about 310 nm.

We checked for the differences in spectral flux between an
approximated Planck function and real spectral data using the Sun and
AD Leo as examples.  We found that the differences can be up to 20-40
\% in the visible and near IR and very large in the far UV. Thus, real stellar input
spectra should be used whenever possible, especially for detailed
chemistry calculations.  However, for the parametric study performed
here the Planck approximation is acceptable for quiet M-dwarfs.


\begin{table*}
\centering
\caption{Stellar parameters used in this study.} \label{star}

\begin{tabular}{c c c c c}\hline\hline
Stellar type  & T$\mathrm{_{eff}}$/K & R/R$_{\odot}$ & M/M$_{\odot}$ & Orbital distance\tablefootmark{a}/AU\\\hline

G2V (Sun)&5770& 1  & 1  &  1\\

M4.5V (AD Leo)\tablefootmark{c}&3400& 0.41 & 0.45  & 0.153\\

M7\tablefootmark{b} &2500 & 0.12 & 0.09   & 0.022\\

M6\tablefootmark{b} &2600 & 0.15 & 0.10  & 0.030\\

M5\tablefootmark{b} &2800 & 0.20 & 0.14   & 0.047\\

M4\tablefootmark{b} &3100 & 0.26 & 0.20   & 0.075\\

M3\tablefootmark{b} &3250 & 0.39 & 0.36   & 0.123\\

M2\tablefootmark{b} &3400 & 0.44 & 0.44   & 0.152\\

M1\tablefootmark{b} &3600 & 0.49 & 0.49   & 0.190\\

M0\tablefootmark{b} &3800 & 0.62 & 0.60   & 0.268\\\hline
\end{tabular}
\hfill\\
\tablefoottext{a}{here the stellar constant equals the solar constant, hence
the same total energy input at TOA is ensured}\\
\tablefoottext{b}{Values from \citet{Kaltenegger2009}}
\tablefoottext{c}{Values from \citet{Leggett1996}}
\end{table*}

In our study, planetary gravity was varied from 1g to 3g (9.8 to 29.4
m s$^{-2}$). Hence, our 1g and 3g cases correspond to 1 
and 11.4 Earth masses and 1
and 1.95 Earth radii \citep{Sotin2007}, respectively. The surface
pressure was kept fixed at 1 bar. As in previous studies
  (e.g. \citealt{Segura2003}, \citealt{Segura2005}), we set the model
  parameters surface albedo and fluxes of chemical species in a way that the mean surface temperature of 288 K
    and observed surface concentrations of the chemical species H$_2$, CO,
    CH$_4$, N$_2$O and CH$_3$Cl are reproduced for present Earth. For the other model scenarios we kept the surface albedo and fluxes constant and calculated the atmospheric concentrations and surface temperature. However, in
  contrast to previous studies, we then placed the modelled extra-solar
  planets at an orbital distance from their central star where their total
  stellar energy input equals that of 1 solar constant (1366~W/m$^2$). We chose this
  approach so that habitable surface temperatures were not enforced 
  and allowed the temperature to adjust in a consistent way to the different
  spectral energy distribution of the M-dwarf stars. Thus, we separated in
  our investigation the effects of the different spectral energy
  distributions from variations of the total incoming flux. We will show in Sect. 4 that the resulting surface temperatures are still
  in the habitable range.

At present, the atmosphere of a terrestrial extrasolar planet has not
been detected directly. Assumptions on their composition are therefore
more or less arbitrary, and are based on our knowledge of the Solar
System. In terms of searching for biomarkers, Earth is the only
reference case we know of a planet with a biosphere. In this
  study, we therefore assume the same initial modern Earth
  atmospheric composition and biomass emissions for the simulated extra-solar planets. The
  final atmospheric composition for the star-planet-scenarios studied
  was then consistently calculated according to the coupled
  climate-chemistry model. Although the
  assumption of an Earth-like biosphere for planets orbiting stars
  that are quite different from our Sun is probably unrealistic, it nevertheless
  serves as a first step towards understanding the effects of
  different environmental factors on habitable planets and their
  atmospheres.

We used specifically the following surface boundary conditions
  to reproduce mean Earth conditions in our Sun-Earth
  reference run. Molecular hydrogen was removed at the surface with a
  constant deposition velocity of 7.7$\times10^{-4}$
  cm~s$^{-1}$. Biomarkers (N$_2$O and CH$_3$Cl) and related
  compounds (CO and CH$_4$) were emitted with constant upwards
  fluxes per surface area in a procedure outlined e.g. in \citet{Grenfell2007}. For example
  a CH$_4$ flux of 531 Tg/year is needed to reproduce modern Earth CH$_4$ surface volume mixing ratios
  and is kept constant for all other simulated
  scenarios. This approach is different to the one employed by \citet{Segura2005}, who
  kept the CH$_4$ surface mixing ratio for the quiet M-star
    runs constant. Furthermore we did not need to change the CO surface
    boundary condition from a constant flux to a constant deposition
    velocity as was done in \citet{Segura2005} for their quiet M-star
    scenarios. As a result, our surface fluxes differ largely from those of 
    \citet{Segura2005} but compare better with the observed values on
    Earth. Again we chose our approach to vary input parameters
  step-by-step, not to mix the influence of different effects in our
  analysis.

To validate our approach, we performed SNR calculations for several
modern Earth transmission features, which were also performed by
\citet{Kaltenegger2009} (their Table 2, single transit events). When we
strictly followed their approach (fixed the atmospheric T-p profiles and
composition to modern Earth), we were able to reproduce their values
to within 2\%.  A direct comparison of the model results with observations
of the Earth transmission spectra by e.g. \citet{Palle2010} is
unfortunately not possible.  The measured spectrum is shortwards of
2.4 $\mu$m and thus overlaps only little with our modelling
range. Furthermore, because our model is stationary and cloud-free a
comparison with real data of a cloudy Earth is difficult and no direct
agreement is expected. A comparison with \citet{Segura2005} is discussed in more detail in Sect. 5.

\section{SNR calculations}
\label{descrsnr}
We here consider emission and transmission spectroscopy in
the near- to mid-IR wavelength range.  We account for photon noise
only to give an upper limit on the detection feasibility.  Additional
noise sources, e.g. instrumental or astrophysical, like zodiacal
light, will further decrease the signal-to-noise ratio (SNR) computed
here. This section describes the method used to calculate the
respective SNR.

\subsection{Emission}
The planetary emission spectrum, $F_p$, is obtained by combining
fluxes (with equal integration times) observed during ($F_1$) and
outside ($F_2$) secondary eclipse
\begin{equation}\label{planetary_signal}
    F_p=F_2-F_1,
\end{equation}
\noindent where $F_1=F_s$ and $F_2=F_s+F_p$, and $F_s$ represents the stellar flux.
The SNR for the planetary spectrum is therefore
\begin{equation}\label{snr_planetary}
    \mathrm{SNR}_p=\frac{F_p}{\sqrt{2\sigma_s^2+\sigma_p^2}}.
\end{equation}
\noindent
Here, $\sigma_s$ and $\sigma_p$ denote the stellar and
  planetary flux standard deviation, respectively.  Assuming
a photon-limited instrument (hence, only Poisson noise from star and
planet) and $F_s\gg F_p$, we then reduce (2) to
\begin{equation}\label{simplified_plan_snr}
  \mathrm{SNR}_p=\frac{F_p}{\sqrt{2\sigma_s^2}}=C_E\cdot\frac{F_s}{\sqrt{2\sigma_s^2}}=\mathrm{SNR}_s\cdot\frac{C_E}{\sqrt{2}},
\end{equation}
\noindent
with $\mathrm{SNR}_s$ the stellar SNR and $C_E$ the contrast between
planetary and stellar signal in the secondary eclipse emission
spectrum. The resulting SNR is conservative compared with the
continuum flux region.

$C_E$ is calculated from the integrated fluxes of a hemisphere as

\begin{equation}\label{emissioncontrast}
  C_E=\frac{F_p}{F_s}=\frac{R_p^2}{R_S^2}\frac{I_P}{I_S},
\end{equation}
\noindent
where $R_P$ and $R_S$ are planetary and stellar radius, respectively,
and $I_P$ and $I_S$ are the spectral energy fluxes (in W~m$^{-2}$~$\mu$m$^{-1}$)
from planet and star, respectively. Here, the star is taken as a
blackbody with a fixed effective temperature, consistent with previous
studies (e.g., \citealp{Kaltenegger2009}). The observable stellar
signal $S_S$ (in J~$\mu$m$^{-1}$) is calculated by

\begin{equation}\label{stellarsignal}
  S_{S}=\frac{R_S^2}{d^2} \cdot I_S \cdot A \cdot t=S_{\mathrm{star}} \cdot A \cdot t,
\end{equation}
\noindent
where $d$ is the distance to the star, $A$ is the telescope area
(assuming a 6.5 m JWST-like aperture in this case) and $t$ the integration
time. To convert into the number of photons, we divide $S_S$ by
$N=h\frac{c}{\lambda}$ ($h$ Planck constant, $c$ speed of light,
$\lambda$ wavelength). From there, we can calculate the spectrally
integrated stellar signal $S_I$ (number of photons)

\begin{equation}\label{}
    S_{I}=\frac{S_{S}}{N} \cdot \frac{\lambda_C}{R},
\end{equation}
\noindent
with $\lambda_C$ the central wavelength and $R$ the specified spectral
resolution.

$S_{det}$, the signal received from the star by the detector, is
calculated as 

\begin{equation}\label{stellarsignal}
    S_{det}=S_{I} \cdot q,
\end{equation}
\noindent
where $q$ is a factor to account for a non-ideal detector and the
overall instrument throughput (assumed for JWST: $q$=0.15
\citealt{Kaltenegger2009}).

The signal-to-noise ratio of the star, $\mathrm{SNR}_s$, is calculated
without taking into account instrumental or read-out noise.

\begin{equation}\label{snrdef}
  \mathrm{{SNR}_s}=\frac{S_{det}}{\sigma_s}=\sqrt{S_{det}}=\sqrt{S_{\mathrm{star}} \cdot A \cdot t \cdot \frac{\lambda_C}{N \cdot R} \cdot q}.
\end{equation}

The SNR values are calculated for integration times equal to the
transit times of considered scenarios ($t=T_D$). The duration, $T_D$,
of a transit is calculated by (assuming that $R_P\ll R_S$,
$i=90^{\circ}$)

\begin{equation}\label{transdur}
    T_D=\frac{P}{2\pi} \cdot 2 \arcsin \left(\frac{R_S}{a}\right),
\end{equation}
\noindent
where $P$ is the orbital period and $a$ the planetary orbital
distance.

From Eq.~\ref{simplified_plan_snr} we then calculate the planetary
SNR ($\mathrm{SNR}_p$).  \newline

Integration times $t_i$ for a certain specified $\mathrm{SNR}_{t_i}$ are
obtained from
\begin{equation}\label{snr_scaling}
    t_i=T_D\cdot \left(\frac{\mathrm{SNR}_{t_i}}{\mathrm{SNR}_p}\right)^2
\end{equation}
\noindent
using the $\sqrt{t}$ scaling of SNR with integration time.
\newline

Because the SNR scales with $\sqrt{F}$ and therefore $d^{-1}$, the integration
times calculated for a specific distance $d_1$  can be easily
scaled to any distance $d_2$:
\begin{equation}\label{distancescaling}
    t(d_2)=t(d_1)\cdot \left(\frac{d_2}{d_1}\right)^2.
\end{equation}

\subsection{Transmission}
The transmission contrast is, to a first order approximation, the
photometric transit depth, i.e.

\begin{equation}\label{transmissioncontrast}
  C_T=\frac{R_P^2}{R_S^2},
\end{equation}

which is much higher than $C_E$ from Eq.~\ref{emissioncontrast}.

In our transmission calculations, we simulated 62 adjacent tangent
height rays through the atmosphere, with a diameter corresponding to
the altitude layer difference in the chemical module. The central
altitudes of the beams correspond to the layer altitudes of our model
atmosphere.  From there, the effective height $h(\lambda)$ of the
atmosphere is calculated as 

\begin{equation}\label{effheight}
    h(\lambda)=\sum_i \left(1-T_i(\lambda)\right)\Delta h_i,
\end{equation}

where $T_i$ is the transmission for ray $i$ at wavelength $\lambda$
and $\Delta h_i$ the altitude difference between consecutive tangent
heights. The additional transit depth, $f_A$, provided by the
atmosphere is therefore obtained by

\begin{equation}\label{addheight}
    f_A(\lambda)=C_T \cdot\left( \frac{\left(R_P+h(\lambda)\right)^2}{R_P^2}-1\right).
\end{equation}

The overall transmission, $T$, of a beam at a given wavelength $\lambda$
is calculated by
\begin{equation}\label{meantransmission}
    T=\frac{1}{H}\sum_i T_i \Delta h_i,
\end{equation}
\noindent
where $H$ is the total height of the  considered atmosphere. Because our
beams cross the atmosphere in equidistant grid points, $T$ is the
simple arithmetic mean of all $T_i$.

The SNR of a spectral feature in transmission is then calculated by
(analogous to Eq.~\ref{simplified_plan_snr})

\begin{equation}\label{transsnr}
   \ \mathrm{SNR}_T=\mathrm{{SNR}_s} \cdot \frac{f_A}{\sqrt{2}}.
\end{equation}

Note the additional factor $\frac{1}{\sqrt{2}}$ in our SNR compared with
\citet{Kaltenegger2009}. The factor results from our approach
that considers the planet signal as resulting from the difference of the
measured stellar flux in and out of transit.

Required integration times or integration times vs. distance are then
obtained by the same scaling as in Eqs.~\ref{snr_scaling} and
\ref{distancescaling}.

\section{Results}

\subsection{Atmospheric profiles}

\subsubsection{Temperature profiles}

\begin{figure}[!h]
\centering  
\includegraphics[width=9.cm]{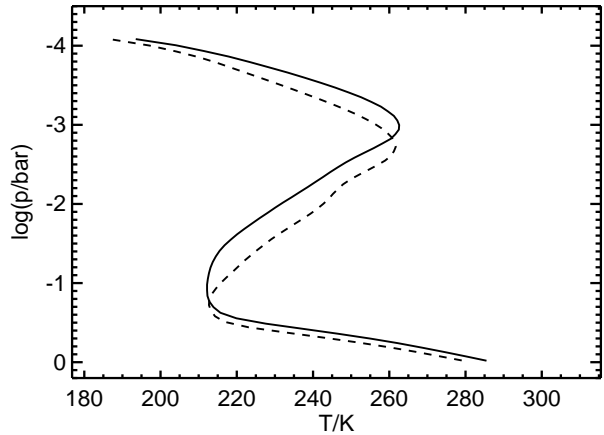} 
\caption{Influence of gravity: Earth-sized 
  (solid) vs. 3g super-Earth planet (dashed) around the
  Sun. } \label{t-p-profile-g}
\end{figure}

Figure \ref{t-p-profile-g} shows the temperature profiles for two
different planet gravity values (1g, 3g) with solar radiation in both cases. The temperature profile for
the modern Earth control run shows a convective troposphere, and the
temperature inversion and  stratospheric temperature maximum caused by
 ozone heating.

The surface temperature decreases for the 3g scenario. This is
because we held the surface pressure constant, which consequently lowered the
vertical column mass.  Less mass means less absorption and therefore
less greenhouse effect, hence lower surface temperatures. Furthermore,
effective ozone heating occurs at higher pressures, $p$, and the
temperature maximum in the stratosphere accordingly decreases to higher $p$ when
increasing gravity. Overall increased stratospheric ozone levels (see
Sect. 4.1.2) lead to enhanced heating in these parts of the
atmosphere.

\begin{figure}[!h]
\centering  
\includegraphics[width=9.cm]{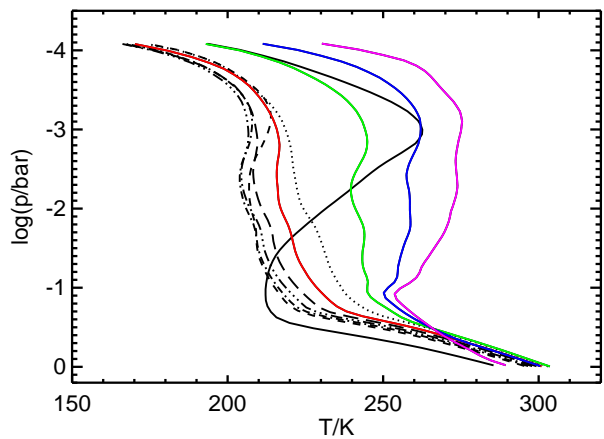} \caption{Influence of
  stellar spectrum: Earth-sized planets around the Sun (black, solid),
  AD Leo (dotted), M0 (dashed), M1 (dash-dot), M2 (dash-dot-dot-dot),
  M3 (long dashes), M4 (red), M5 (green), M6 (blue), and M7-type stars
  (magenta).}\label{t-p-profile-star}
\end{figure}

Figure \ref{t-p-profile-star} shows the influence of the stellar type on the
temperature profile for the Earth control case as well as 1g planets
around AD Leo and the M0 to M7 stars from Table \ref{star}.
Because unlike in previous approaches, we did not fix the
  surface temperature of our modelled planets by different stellar insolations at TOA, it is interesting to
  note that although Earth-like planets around M-dwarfs show somewhat
  increased surface temperatures, they always remain well within the
  habitable range for a total received energy flux of 1 solar
  constant. However, surface temperatures for the 1g scenarios are up to 20 K higher than
  for the Earth-Sun case. For the super-Earth scenarios similar changes occur. Here, surface temperatures are higher for the M-type central stars than for the Sun, and increase from M0 to M4-type stars (not shown). The increase occurs because Rayleigh scattering contributes less
    to the planetary albedo owing to the $\lambda^{-4}$-dependence of the
    Rayleigh scattering coefficient.  In addition, we find enhanced
greenhouse warming (see below).  Overall, however, the differences
  in spectral energy distribution alone have only a moderate effect on
  mean surface temperatures in the context of habitability in general. 
  They may nevertheless significantly alter the planetary environment
   when studying the detailed atmospheric conditions (for example in
  a 3D approach that includes dynamical effects such as tidal-locking (see e.g. \citealt{Joshi1997})).

The main differences between the Sun and the M-star planets are
  found in the stratosphere, as already noted by \citet{Segura2005}.
  The atmospheres of the modelled M-star planets do not show a
  stratospheric temperature inversion up to type M5 (M4) for the Earth-sized (super-Earths) planets. This is caused by
  the different stellar energy flux distribution of M-dwarfs compared
  with the Sun.  The reduced UV flux longwards of about 250 nm results in
  less heating by ozone and therefore a reduced stratospheric
  inversion. We confirm this finding by \citet{Segura2005} for the
  active star AD Leo.  In addition, our study also includes model runs for
  several quiet M-dwarfs with otherwise the same parameters for their
  orbiting planets.  As can be seen in Fig. \ref{t-p-profile-star},
  the planetary stratospheric temperatures increase from hot (M0) to
  cooler M central stars (M7), even building up a stratospheric
  inversion again for M6 and M7 stars. This is caused by increasing
  amounts of methane and water in the mid-atmosphere (see following
  section) which lead to enhanced absorption and therefore heating of
  the stratosphere.

As already noted by \citet{Segura2003} and \citet{vparis2008}, the
RRTM radiation scheme as applied in the model has been
constructed for atmospheres that do not deviate too much from Earth
conditions both in terms of temperature structure and composition
\citep{Mlawer1997}. When applied to atmospheres that significantly deviate
from modern Earth, it is important to check that the
atmospheric T-p-profiles do not leave the validity range of RRTM. If
this is the case, the resulting effect on the heating and cooling rates must be
estimated.

We compared the computed T-p-profiles for the atmosphere considered
here to the validity range of RRTM.  We found that T-p-profiles for
most M-dwarf central stars leave the validity range significantly only
in the upper atmosphere. As shown by \citet{vparis2008},
variations in upper stratospheric optical depths have a negligible
influence on lower atmospheric temperature profiles. Most of the
emission spectra originate in pressures higher than 0.1 bar,
except for planets around very cool M dwarfs. 
For a given chemical composition, the transmission
    spectra are nearly independent of T-p structure (in contrast to
    the emission spectra), hence uncertainties in the IR radiative
    transfer of the climate model are unlikely to significantly influence  the
    transmission spectra. To investigate whether these temperature errors
    might in turn influence  the chemical composition of the atmosphere
    by means of e.g.~temperature-dependent reaction rates, additional tests
    were performed in which the stratospheric temperatures were
    arbitrarily changed by $\pm$10K.  Observed changes in
    concentrations and column amounts did not exceed 5-20\% depending
    on species, with methane showing the largest effect. However, no difference could be discerned in
    the spectra.  Hence, the
influence of errors in stratospheric optical depths on the planetary
spectra is likely to be small.  A comparison of the thermal fluxes of
RRTM with thermal fluxes computed by the line-by-line radiative
transfer code SQuIRRL showed an agreement of the total outgoing fluxes
within 5.5\%.

The surface temperature decreases again for the 1g
  planet that orbits the M7 star. In view of the above discussion on
  model validation, we consider the M7 case to be at the limit of the
  application range. We did not study cooler central stars here,
  because the results would be out of the application range over
  most of the atmosphere. For the same reason, we also exclude the M6 and M7 stars from our further analysis of 3g super-Earth scenarios.
  Nevertheless, the M7 case indicates an
  interesting trend.  Increased absorption of stellar radiation in
  mid-atmosphere levels owing to massively increased H$_2$O and CH$_4$
  concentrations results in less radiation reaching the
  ground. Whether this trend would continue for planets that orbit even
  cooler M dwarf stars would need to be further investigated with a model
  adapted to these conditions.

\subsubsection{Chemical profiles}\label{chemicalprofiles}

Ozone, nitrous oxide, methyl chloride, methane, and water are
the major biomarker molecules and related compounds. In
this section we give a broad overview of the
    concentrations of O$_3$, N$_2$O, CH$_3$Cl, CH$_4$ and H$_2$O. A detailed
analysis of the chemical processes for the model runs presented here
will be given in a forthcoming publication
  (\citet{Grenfell2010}, hereafter paper II).

\begin{figure}[!h]
\centering  
\includegraphics[width=9.cm]{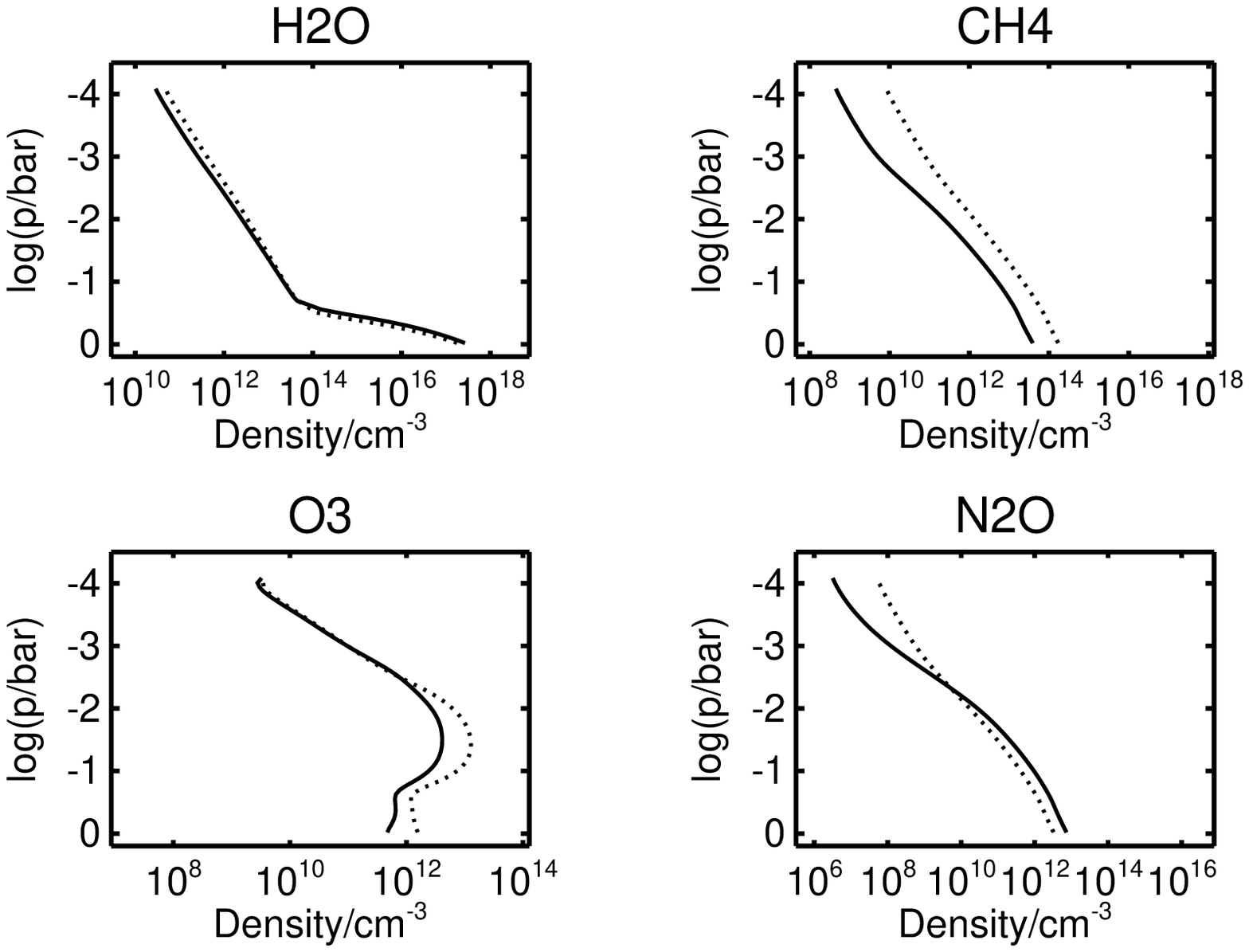} 
\caption{Chemical profiles for an
Earth-sized planet (solid) and a} 3g
  super-Earth (dotted) around the Sun.\label{compgrav_chem}
\end{figure}

\paragraph{Chemical profiles for 1g and 3g scenarios}

\hfill\\   
Because of the decrease in total column mass for the 3g planetary scenario compared with the 1g case (because of the fixed surface pressure of 1 bar), the decrease in greenhouse effect (GHE) resulted in lower tropospheric temperatures. This in turn led to a small lowering in the tropospheric H$_2$O number densities, as can be seen in the upper left panel of Figure~\ref{compgrav_chem}. Although the impact on the number density is small, the impact on the H$_2$O column is large (it is five times lower for the 3g case than for the 1g scenario) owing to the overall decrease in column mass. The decrease in the H$_2$O column led to a lowering in OH to a surface number density of 1.24$\cdot$10$^6$ molec.~cm$^{-3}$ compared with 1.31$\cdot$10$^6$ molec.~cm$^{-3}$ for the 1g case. Because OH is a major sink for CH$_4$, its number density and column increased. While the surface CH$_4$ number density increased from 3.9$\cdot$10$^{13}$ molec.~cm$^{-3}$ (1g) to 1.7$\cdot$10$^{14}$ molec.~cm$^{-3}$ (3g) (see upper right panel in Figure~\ref{compgrav_chem}), the column increased by 44\% only, again owing to the lower total column mass for the 3g scenario.

The O$_3$ number density increased for the 3g planetary scenario (see lower left panel in Figure 4)
in the lower to mid-atmosphere compared with the 1g case. The overall smaller column amount at 3g led to
a small strengthening of the Chapman source at these levels (for more details see paper II). The main ozone
photochemical features typical of the Earth were maintained at 1g and 3g, e.g. in both cases HOx and NOx
catalytic cycles were important in the upper and lower atmospheres respectively, whereas the smog cycle
was important for ozone production in the troposphere for both 1g and 3g.
Compared with changing the stellar type (see below), the effects of increasing gravity are rather small, however. Nevertheless, a more detailed
  study of the chemical effects in simulated super-Earth planet
  atmospheres is given in paper II.

\begin{figure}[!h]
\centering  
\includegraphics[width=9.cm]{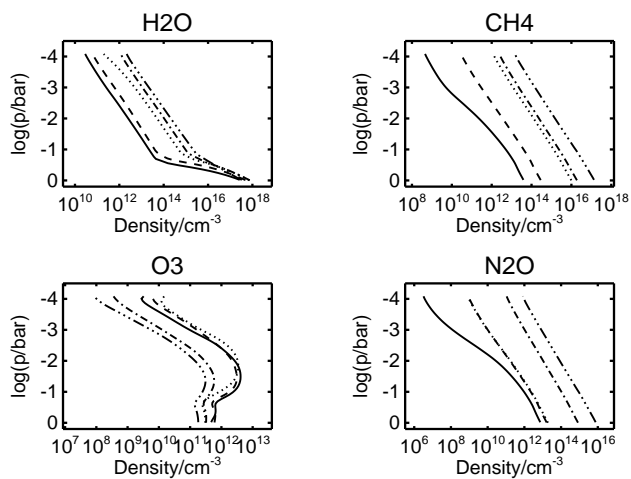}
\caption{Influence of central star type on chemical atmospheric profiles of an Earth-sized planet: Sun (solid), M0 (long dashed), AD Leo (dotted), M5 (dot-dashed), M7 (triple-dot-dashed).}\label{compstar_chem}
\end{figure}

\paragraph{Chemical profiles for different stellar types}
\hfill\\
The response of biomarkers and important greenhouse gases 
   to different central stars is
  shown in Fig. \ref{compstar_chem} for some examples. Increasing the
  M-dwarf stellar type and consequently decreasing the amount of UV
    radiation (except for the active star AD Leo) has a significant
  effect on the photochemical responses in the atmosphere. We 
  briefly discuss the main photochemical responses. For more details we refer
  to the analysis of the chemical processes made in paper II. 

We first note a strong increase in planetary stratospheric
  water content when moving from the Sun via M0 stars to our coolest
  central star, M7. Furthermore, methane  significantly increases for cooler central stars.
One cause for increased stratospheric water is the increase in tropospheric water abundance owing to higher surface temperatures.  Another reason is the reduced UV flux of quiet M dwarfs compared with the Sun, which leads to less photolytic destruction
  (see Fig. \ref{inputspectra}). As a result, the abundance
  of OH is also reduced, which can lead to reduced  CH$_4$ destruction. On the other hand, 
increased methane abundances can lead to faster CH$_4$ oxidation and therefore to increased stratospheric water production,
as suggested by \citet{Segura2005}. Detailed investigations show that other catalytic cycles
also play a role. This analysis is presented in paper II.

\citet{Segura2005} pointed out that the
  photolytic destruction of methane in Earth-like, high-oxygen planets
  is slow, and in combination with drastically reduced sinks by OH
  this could lead to a "methane runaway". Possible
  methane runaway feedbacks are quite well investigated for the Earth's
  atmosphere (e.g. \citealt{Prather1996}). The main feedback is
  positive, where more methane leads to less OH, therefore slower
  methane in-situ oxidation, which favours  methane even more. There is
  also an opposing, negative feedback on Earth, e.g. methane warming
  leads to a damper atmosphere, which favours an increase in
  OH. There exist potentially complex interactions
    e.g. with NOx emissions, where increased NOx generally favours fast
    partitioning of HO$_2$ into OH (via: NO + HO$_2$ $\rightarrow$ OH
    + NO$_2$) which in turn affects the CH$_4$ runaway. \citet{Lelieveld2002}
    provided an overview of these effects. In model simulations as
  presented here and in \citet{Segura2005}, methane runaway can occur
  as an unphysical scenario for quiet M-dwarfs if the assumed methane
  fluxes exceed the chemical sinks and loss processes in the model. Our methane fluxes are lower than those of \citet{Segura2005}, and
  this methane runaway does not occur for our scenarios.

\begin{figure*}
 \centering
\includegraphics[width=18cm]{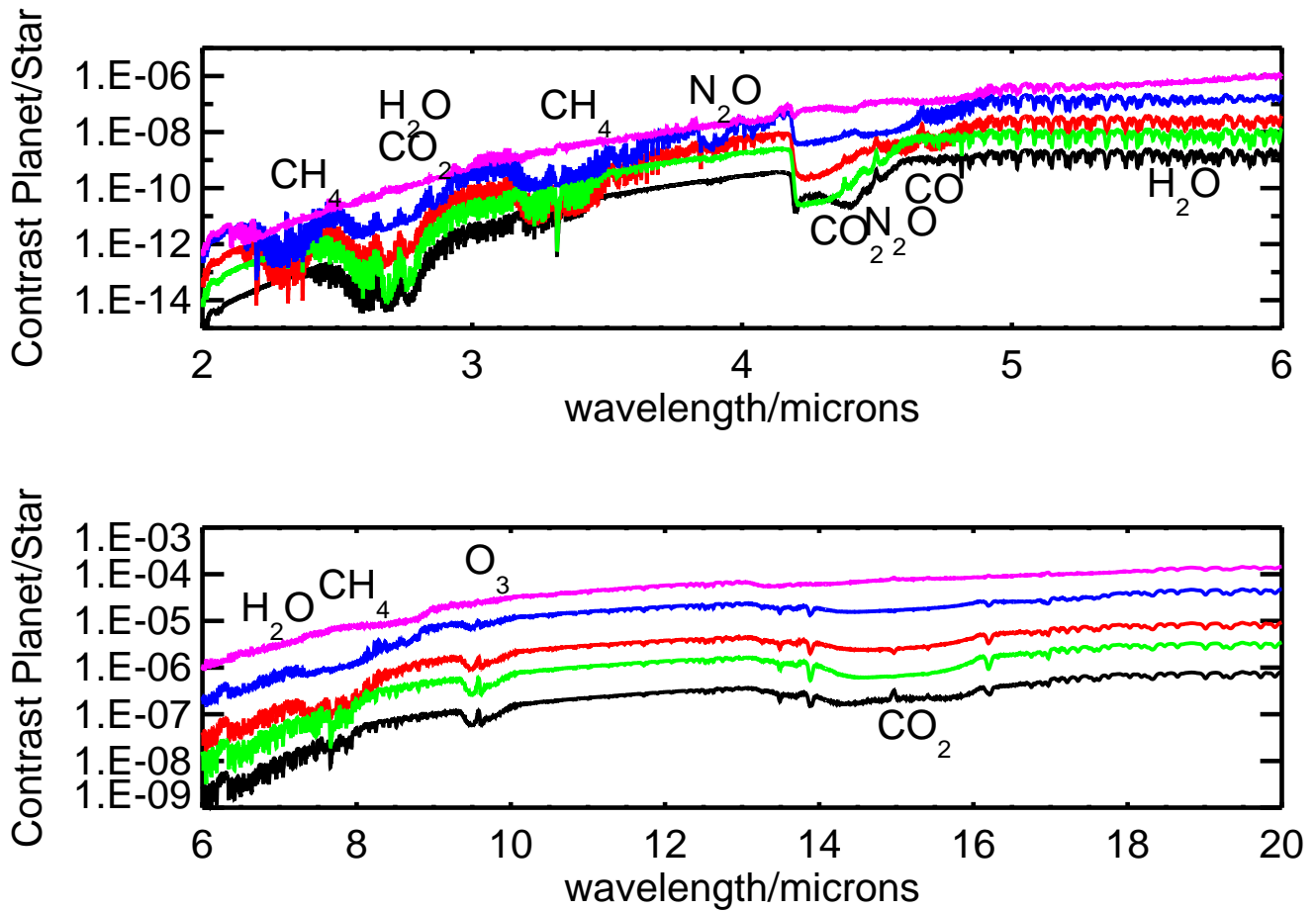}
\caption{Contrast spectra for an Earth-sized planet (1 g) around the
  Sun (black), AD Leo (red), M0 (green), M5 (blue), and M7
  (magenta).}\label{emis_1g_1gm}
\end{figure*}

Fig. \ref{compstar_chem} shows the ozone profiles for the Earth-sized planet
  scenarios.  The ozone profile of the Earth-Sun reference run
  and the planets that orbit AD Leo are similar, whereas ozone decreases
  for cooler M-dwarfs and is significantly reduced in the M4 to M7
  cases. Ozone abundances are controlled by the balance between
  photochemical sources via the established Chapman and smog
  mechanisms and the catalytic O$_3$ sinks, which depend on
  concentrations of e.g. hydrogen and nitrogen-oxides. A key
  result of the analysis of chemical processes in paper II is that
  ozone production changes from being dominated by the Chapman
    to the smog mechanism with increasing stellar type, thus with
  decreasing effective temperature and UV flux. The weaker UV
  flux from the cooler M dwarfs leads to slower photolysis of the O$_2$ molecule, which is required by
  the Chapman mechanism. Smog production is still slow in
  comparison. This explains the decline of ozone for cool, quiet M-dwarf central stars.  The simulated planet that orbit the active star
  AD Leo though shows a pronounced ozone layer similar to the
  Earth owing to its higher UV flux compared with the quiet M4 and
  M5 stars. The prominent ozone layer for active M dwarfs was also
  found by \citet{Segura2005}.  More details on the changing ozone
  production mechanisms for cool central stars are explained in paper
  II. 

The changing stellar flux of M dwarfs with increasing M-star class also leads to a strong increase
in N$_2$O densities (Fig. \ref{compstar_chem}). This is again caused
by reduced photolysis rates due to the lower UV flux. Again, the
effect is smaller for active M-dwarfs like AD Leo, where the photo\-lysis
rates are much higher.

We also investigated the abundance of the biomarker molecule
  CH$_3$Cl (not shown here), because
   \citet{Segura2005} showed this to potentially cause significant absorption bands for high
  concentrations and active M-dwarf stars. In
  contrast to \citet{Segura2005}, however, our required
  surface CH$_3$Cl flux to reproduce mean Earth conditions is
  a factor of 2 lower, which also better agrees with the
    observed estimates. We find an increase in CH$_3$Cl abundances for
  cooler M dwarfs owing to the reduced UV flux, similar to the increase
  in methane. However, in all cases studied, no significant spectral
  absorption could be seen in the emission spectra, whereas
    small absorption at 13.7 $\mu$m for the coolest central star (M7)
    can be found in the transmission spectrum (see
    Fig. \ref{reltrans_ref_1m}). This molecule was therefore not
  investigated in more detail in this work.

\subsection{Emission spectroscopy}

\subsubsection{Emission spectra}

The IR emission spectra of extra-solar planets can be measured for
transiting planets by observing the secondary eclipse. The planetary
signal is derived from the difference of flux during and
out-of-eclipse. Below, we show the expected flux contrast (planet/stellar flux) for computed planetary IR
emission spectra.

The modern Earth reference contrast spectrum is shown in Fig.
\ref{emis_1g_1gm} (black line). The spectrum follows a blackbody
spectrum with absorption bands of, e.g., water (2.7 $\mu$m, the
broad 6.3 $\mu$m band and the rotation band longwards of 15
$\mu$m), methane (at 3.3 and 7.7 $\mu$m), ozone (9.6 $\mu$m) or
carbon dioxide (2.7, 4.3 and 15 $\mu$m). The contrast between
stellar and planetary flux is very low, reaching $10^{-6}$ in the
mid-IR. Because of the rapid decrease of the planetary Planck function
towards shorter wavelengths, the contrast falls below $10^{-10}$
in the near-IR.

\begin{figure}
\centering
\includegraphics[width=9.cm]{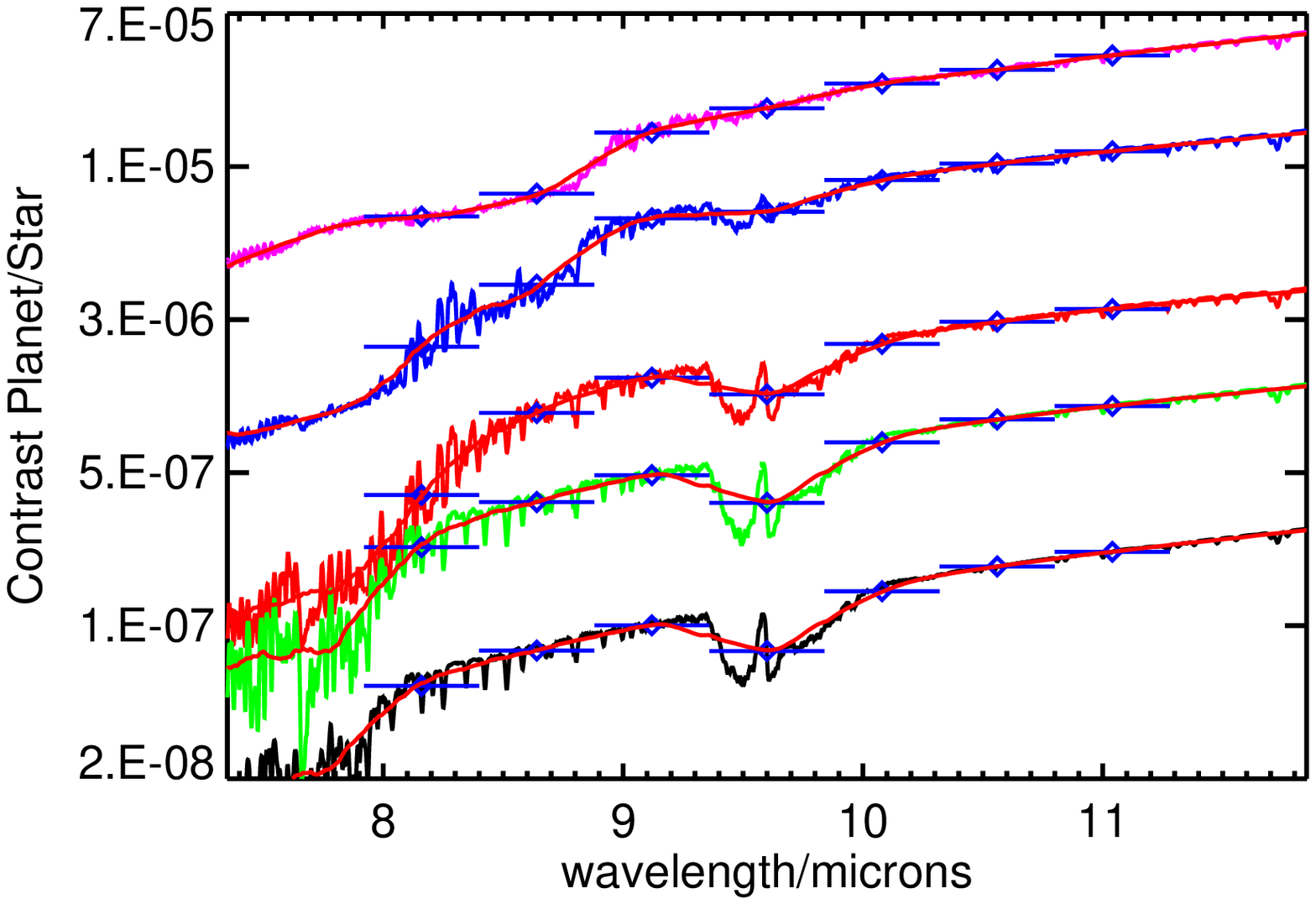}
\includegraphics[width=9.cm]{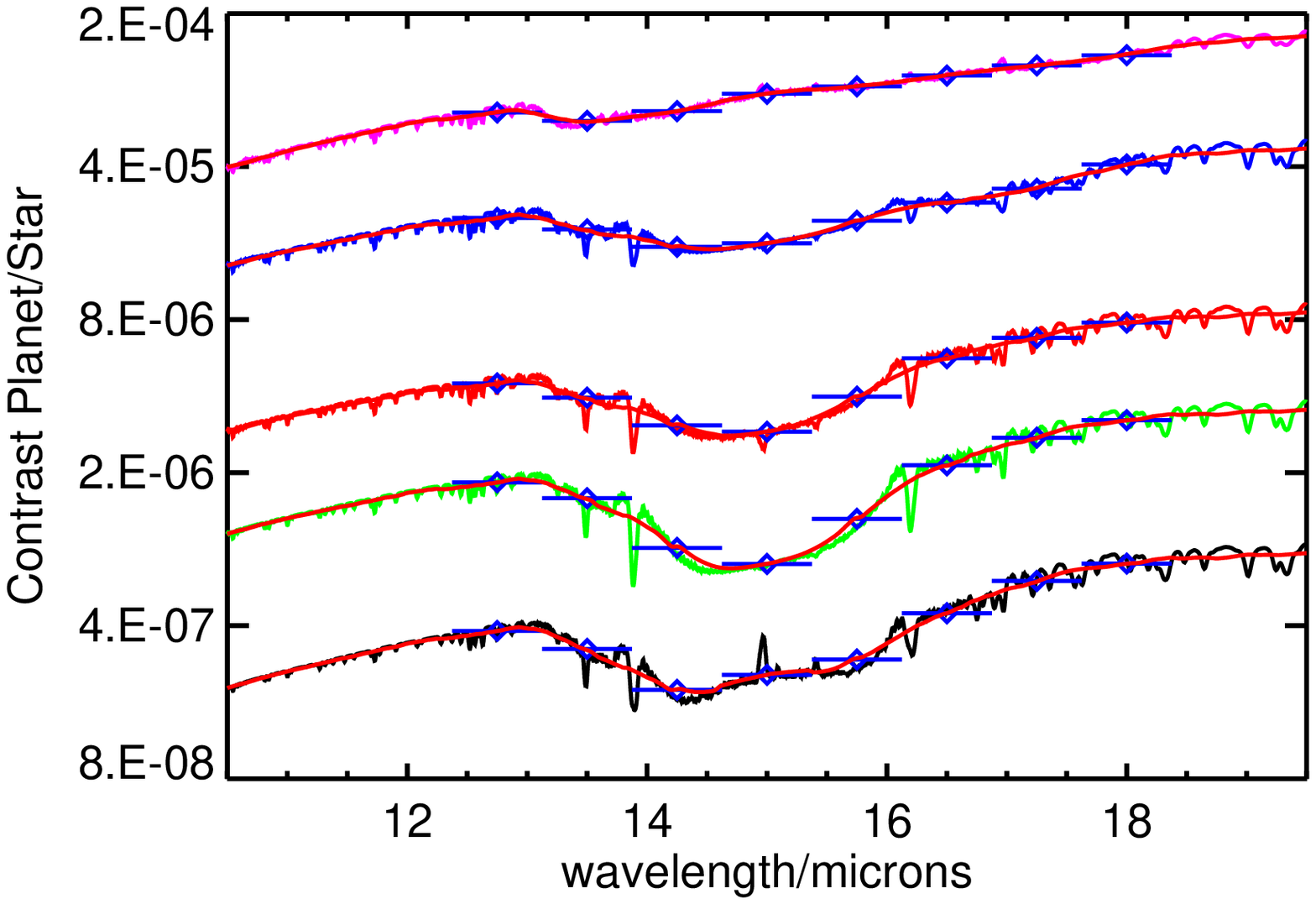}
 \caption{Contrast spectrum of the 9.6 $\mu$m ozone
band and  the 15 $\mu$m carbon dioxide
  band for an Earth-sized planet around the Sun (black), AD Leo (red), M0 (green), M5 (blue), and M7 (magenta).
Blue diamonds: binned to $R$=20, red: smoothed spectra at the same resolution.}\label{co2_earth_zoom}
\end{figure}

\begin{figure}
  \centering 
  \includegraphics[width=9.cm]{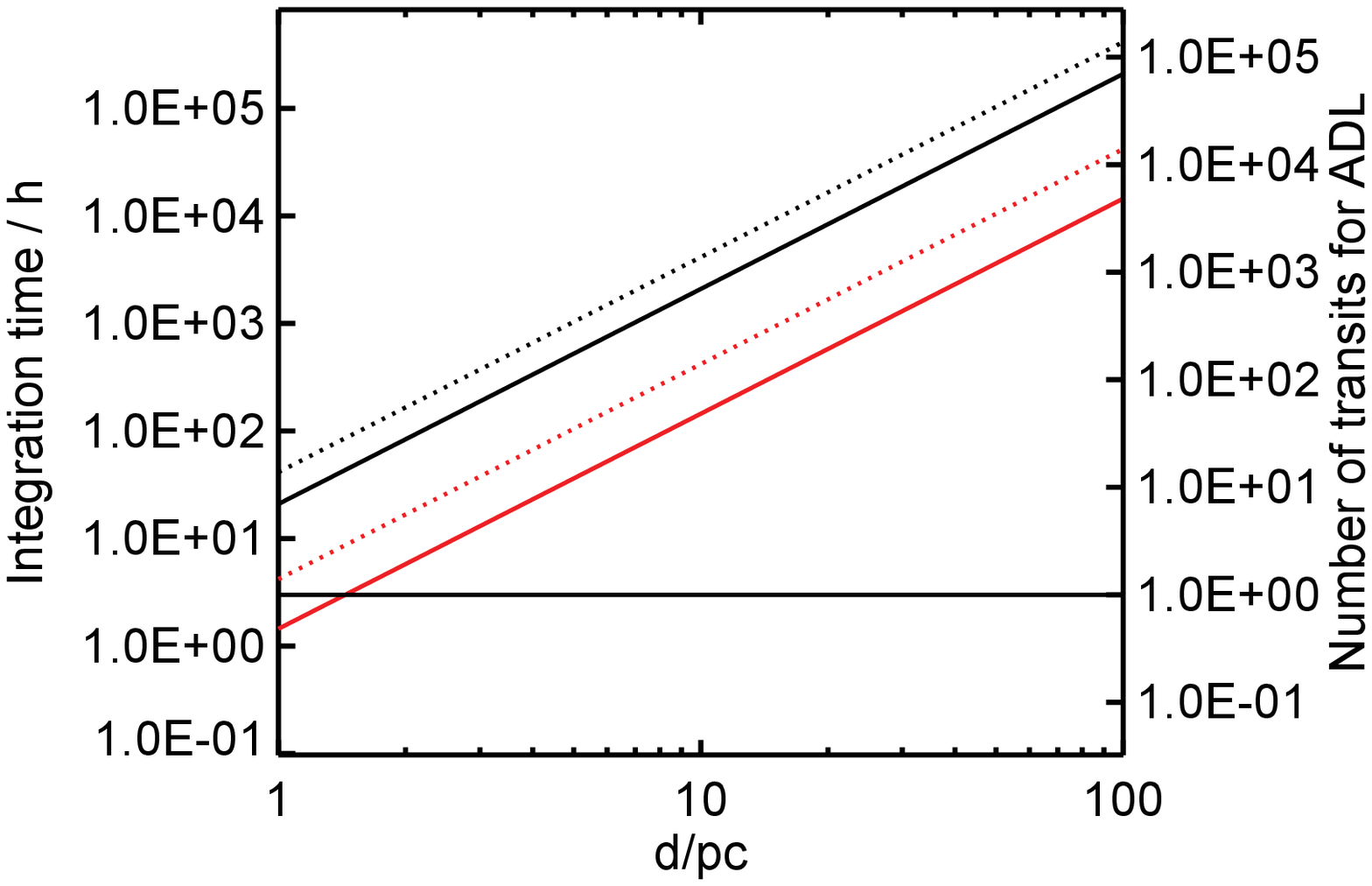}
  \includegraphics[width=9.cm]{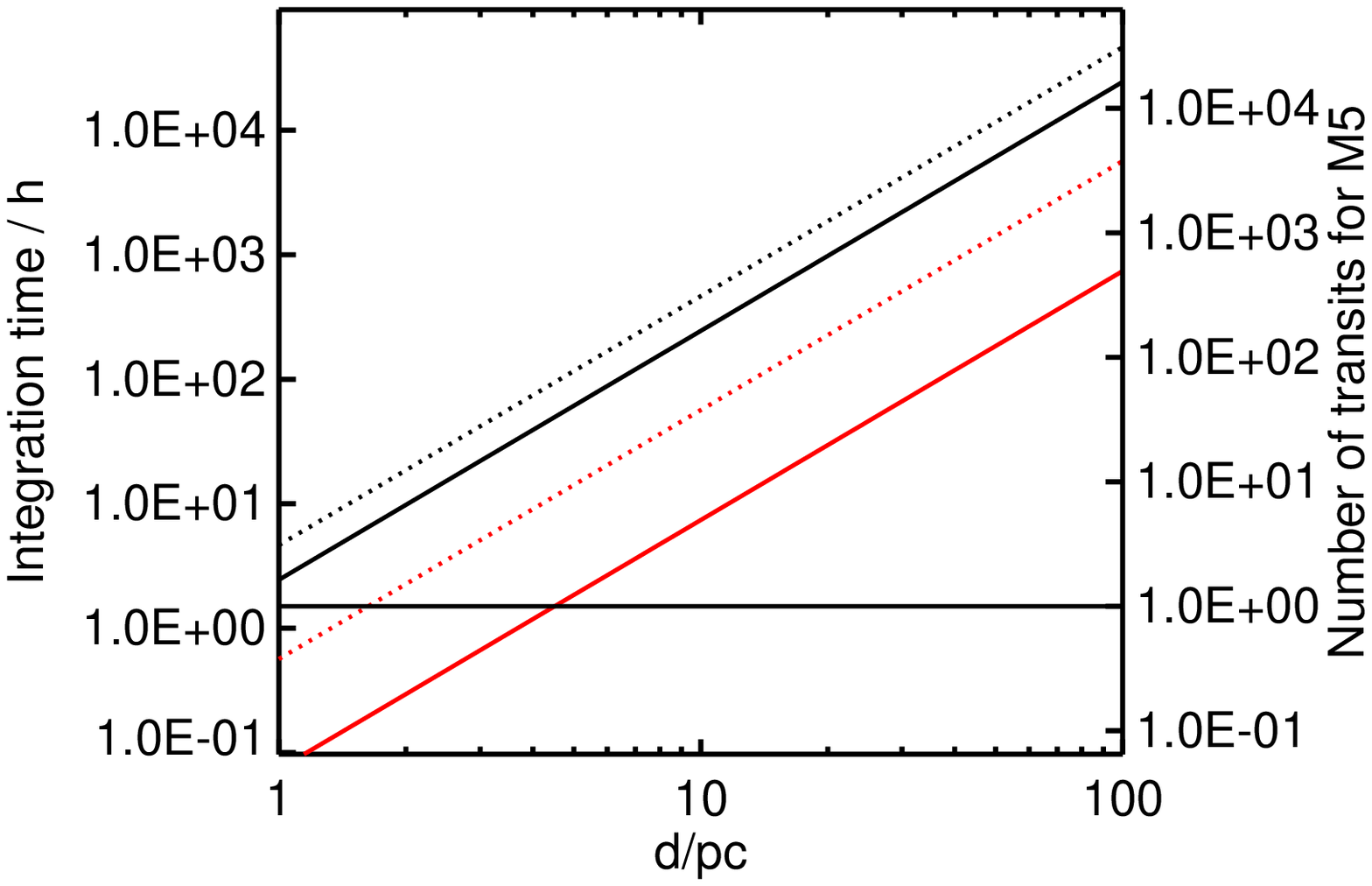}
  \caption{Emission spectra: Integration time in hours needed for SNR=3 as a
    function of stellar distance of the 9.6 $\mu$m (dotted) ozone and
    the 15 $\mu$m (solid) carbon dioxide spectral feature at $R=20$
    for an Earth (black) and a 3g super-Earth (red) around AD Leo (top)
    and an M5 dwarf central star (bottom).  Telescope configuration: JWST.}\label{time_2}
\end{figure}

Figure \ref{emis_1g_1gm} also compares the spectra of an Earth-like
planet (1g) when changing the central star (Sun, AD Leo, M0, M5,
M7). The overall contrast is improved for terrestrial planets around M
dwarfs, as expected. Because of the much lower effective temperatures
and the smaller radius of M-dwarf stars, the contrast increases by
about two orders of magnitude and eventually reaches about $10^{-4}$ in
the mid-IR, which illustrates the improved potential of detecting
planets around M-type dwarf stars.

Concerning molecular absorption bands, noticeable differences to the
Earth-Sun system are found in the 2.3, 3.3 and 7.7
$\mu$m methane bands and the 3.8 and 4.5 $\mu$m band
of nitrous oxide. These molecules show higher concentrations in the
atmosphere of M dwarf planets compared with Earth (see
Fig. \ref{compstar_chem}), therefore we would expect their
spectral signatures to be more
pronounced. Additionally, the 4.7 $\mu$m band of CO
    becomes more prominent owing to its higher concentrations in the
    atmosphere (not shown). This
  can indeed be observed for the planets that orbit the quiet M0-M3 dwarfs
  stars. However, the absorption bands of N$_2$O, CH$_4$ become weaker for cooler M stars and almost
    completely disappear for an Earth-sized planet around the M7 star, despite the greatly enhanced
    abundances. This is caused by the temperature structure of these
    planets (see Fig. \ref{t-p-profile-star}). The emission in the
    CH$_4$ and N$_2$O bands originates in the middle atmosphere because of
     the large concentrations of these gases, i.e. the atmosphere
    becomes transparent only at much higher altitudes than for Earth
    around the Sun. At these altitudes, the temperature is almost as
    high as the surface temperature, hence the contrast between
    continuum (transparent atmosphere) and absorption band is
    reduced. The absorption cannot be discerned from the spectrum any
    longer.  Thus, the atmospheric composition may be difficult to
  detect via emission spectroscopy in planets orbiting quiet and very
  cool M dwarfs, despite a more favourable planet/star flux contrast
  ratio.

Another remarkable difference occurs for the 4.3 and 15 $\mu$m
fundamental bands of carbon dioxide. The absence of a temperature
inversion and the much colder stratosphere for the M0 to M5-type stars (see Fig.
\ref{t-p-profile-star}) have a noticeable effect, as already
  noted by \citet{Segura2005}. The 4.3 $\mu$m band absorption is
stronger for the M0 and AD Leo cases, and in the 15 $\mu$m band the
emission peak is absent. However, as already
    discussed, our model runs for quiet M stars show that for cool
    stars, e.g. the M7 case, the emission spectrum originates mainly
    in the upper atmosphere for almost the entire spectrum (not only
    the CO$_2$ fundamentals), and therefore molecular absorption features
    are weak.

\begin{figure}
\centering  
\includegraphics[scale=0.4]{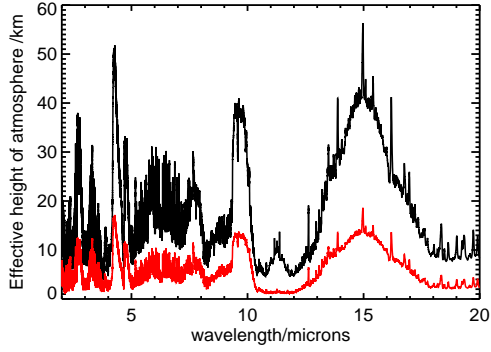}
\caption{Effective
tangent heights: Earth and 3g
 super-Earth (red) around Sun.}\label{height_ref_3g}
\end{figure}

Figure \ref{co2_earth_zoom} shows the 15 $\mu$m carbon dioxide and the
9.6 $\mu$m ozone band at high and low wavelength
resolutions. The CO$_2$ band is clearly visible already at a resolution of
$R=20$, although the emission peak close to the line centre is not
resolved. This emission peak is an indicator of the atmospheric
temperature inversion, hence a potentially important spectral
feature. To resolve the emission peak, however, at least $R\approx 200$
is necessary. The same conclusion holds for other interesting very
narrow features. The ozone absorption at 9.6 $\mu$m is still visible
at $R=20$. For a detection of the molecule, low resolution seems to be
sufficient. We also performed simulations for even lower resolution
($R=5$). In this extreme, however, ozone is barely visible against the
surrounding continuum and becomes difficult to detect.

\subsubsection{Signal-to-noise-ratios}

We  calculated the expected SNR for different planet-star
configurations. The chosen telescope aperture is as for JWST with an
assumed efficiency of 0.15 (\citealp{Kaltenegger2009}).  Integration
times correspond to a single transit event (i.e., 13 hours for
super-Earth around Sun and 3 hours for super-Earth around AD Leo).
For the Sun-Earth case, SNR values (not shown) are always below
0.15. Earth-like planets around M stars also show low SNR values.
Table \ref{summaryemissearth} summarises the achievable SNR values for
a 3g super-Earth around the Sun, AD Leo, M0, and M5.  As can be
seen from Table \ref{summaryemissearth}, the single transit SNR is
below unity for all absorptions, except for low resolution and late M
dwarf types.

\begin{table*}
\centering
\caption{SNR values for emission spectra of a 3g
super-Earth at 10 pc\tablefootmark{a}\label{summaryemissearth}}
\begin{tabular}{c c c c c c c}\hline\hline
Stellar type & Molecule & $\lambda$ /$\mu$m & $R$=5 & $R$=20 & $R$=500 &
$R$=2000\\\hline
Sun& CH$_4$& 3.3& 1.4$\cdot 10^{-4}$ & 5.2$\cdot 10^{-5}$&  9.0$\cdot 10^{-6}$& 4.1$\cdot 10^{-6}$\\
    &   CO$_2$& 4.3& 2.0$\cdot 10^{-3}$&  1.6$\cdot 10^{-4}$&  3.6$\cdot 10^{-5}$& 1.7$\cdot 10^{-5}$\\
    &   CH$_4$& 7.7& 1.2$\cdot 10^{-1}$&  5.5$\cdot 10^{-2}$&  9.7$\cdot 10^{-3}$& 5.8$\cdot 10^{-3}$\\
    &   O$_3$ & 9.6& 3.7$\cdot 10^{-1}$&  1.4$\cdot 10^{-1}$&  2.8$\cdot 10^{-2}$& 1.3$\cdot 10^{-3}$\\
    &   CO$_2$& 15.0& 6.2$\cdot 10^{-1}$&  2.6$\cdot 10^{-1}$&  5.8$\cdot 10^{-2}$& 2.8$\cdot 10^{-2}$\\\hline
AD Leo& CH$_4$& 3.3& 1.4$\cdot 10^{-4}$ & 2.4$\cdot 10^{-5}$&  6.7$\cdot 10^{-6}$&2.7$\cdot 10^{-6}$\\
    &   CO$_2$& 4.3& 3.3$\cdot 10^{-3}$&  2.3$\cdot 10^{-4}$&  3.1$\cdot 10^{-5}$& 1.5$\cdot 10^{-5}$\\
    &   CH$_4$& 7.7& 1.3$\cdot 10^{-1}$& 4.4$\cdot 10^{-2}$ & 6.9$\cdot 10^{-3}$&  3.7$\cdot 10^{-3}$\\
    &   O$_3$ & 9.6& 6.9$\cdot 10^{-1}$& 2.5$\cdot 10^{-1}$ & 5.0$\cdot 10^{-2}$  & 2.4$\cdot 10^{-2}$ \\
    &   CO$_2$& 15.0& 1.12 & 4.3$\cdot 10^{-1}$ & 8.3$\cdot 10^{-2}$ & 4.2$\cdot 10^{-2}$\\\hline
M0& CH$_4$& 3.3& 1.8$\cdot 10^{-4}$ & 4.7$\cdot 10^{-5}$&  9.4$\cdot 10^{-6}$&4.3$\cdot 10^{-6}$\\
    & CO$_2$& 4.3& 2.8$\cdot 10^{-3}$&  9.1$\cdot 10^{-5}$&  1.1$\cdot 10^{-5}$& 5.7$\cdot 10^{-6}$\\
    & CH$_4$& 7.7& 1.3$\cdot 10^{-1}$&  4.3$\cdot 10^{-2}$&  5.3$\cdot 10^{-3}$& 3.2$\cdot 10^{-3}$\\
    & O$_3$ & 9.6& 5.5$\cdot 10^{-1}$&  2.1$\cdot 10^{-1}$&  4.3$\cdot 10^{-2}$& 2.0$\cdot 10^{-2}$\\
    & CO$_2$& 15.0& 8.0$\cdot 10^{-1}$&  2.8$\cdot 10^{-1}$&  5.4$\cdot 10^{-2}$& 2.7$\cdot 10^{-2}$\\\hline

M5& CH$_4$& 3.3  & 3.5$\cdot 10^{-4}$& 1.8$\cdot 10^{-4}$&  3.1$\cdot 10^{-5}$& 1.4$\cdot 10^{-5}$\\
    & CO$_2$& 4.3& 6.4$\cdot 10^{-3}$& 3.6$\cdot 10^{-3}$&  7.9$\cdot 10^{-4}$& 3.8$\cdot 10^{-4}$\\
    & CH$_4$& 7.7& 3.9$\cdot 10^{-1}$& 2.2$\cdot 10^{-1}$&  4.7$\cdot 10^{-2}$& 2.4$\cdot 10^{-2}$\\
    & O$_3$ & 9.6& 9.2$\cdot 10^{-1}$& 4.9$\cdot 10^{-1}$&  9.8$\cdot 10^{-2}$& 4.9$\cdot 10^{-2}$\\
    & CO$_2$&15.0& 2.57           & 1.35                 &  2.8$\cdot 10^{-1}$& 1.4$\cdot 10^{-1}$\\\hline
\end{tabular}
\hfill\\
\tablefoottext{a}{Integration time is a single transit duration.
Telescope configuration: JWST.}\\
\end{table*}

Table \ref{summaryemissearth} shows that emission spectroscopy of
terrestrial planets in the HZ provides very low SNR for a single transit
duration exposure for the 3.3 $\mu$m methane, the 4.3 $\mu$m
carbon dioxide, the 6.3 $\mu$m water and the 7.7 $\mu$m methane band.
However, for the 9.6 $\mu$m band of ozone and the 15 $\mu$m band of
carbon dioxide, SNR values are more encouraging at low resolution.
Nevertheless, several transits need to be added to reach realistic
SNR.

In Fig. \ref{time_2} we show the integration times needed to obtain a
SNR of 3 for $R=20$ as a function of stellar distance. The planetary
scenarios are the 1g Earth-like and 3g super-Earth planet around AD Leo and an M5
dwarf. As examples, we chose the ozone band and the broad carbon
dioxide band at 15 $\mu$m. The integration time is indicated in units
of transit time on the right y-axis. The figure indicates that for 3g
super-Earths a SNR of 3 in the 15 $\mu$m CO$_2$ band could be achieved
in a single transit for the cool M5 dwarf closer than about 10 pc. Higher resolution is probably required for more
detailed characterisation, e.g. for the detection of a temperature inversion.

\subsection{Transmission spectroscopy}
\subsubsection{Transmission spectra}
During primary eclipse the light of the central star passes through
the atmosphere of the transiting planet and absorption by atmospheric
molecules will change the apparent radius of the planet according to
their scale height.  Generally, transits are observed in the optical
wavelength range because of the higher stellar flux and therefore
better SNR. However, M-type stellar radiation is shifted somewhat
to IR wavelengths and, furthermore, many interesting biomarker or
related molecules show absorption in the IR. Below we
therefore investigate the signal for primary transit spectroscopy in
the near- to mid-IR range.

The wavelength-dependent effective tangent height of the planetary
atmosphere blocks the stellar light, hence increases the apparent
radius of the planet. The effective height $h$ (see Eq.~\ref{effheight}) is shown in Fig.
\ref{height_ref_3g} for an Earth-sized planet (black line) and a super-Earth (red)
around the Sun. $h$ is mainly below 20 km, except for the strong
methane 3.3 $\mu$m band, the ozone 9.6 $\mu$m band and the three 2.7,
4.3 and 15 $\mu$m bands of carbon dioxide in the case of the
Earth. The super-Earth scenarios, however, may not be best suited for
transmission spectroscopy, because the effective tangent height is of the
order of the atmospheric scale height, which varies as $g^{-1}$. From
Eq.~\ref{addheight}, the planetary signal decreases as $g^{0.4}$ when
using the mass-radius-relationship from \citet{Sotin2007}. This effect
is illustrated in Fig. \ref{height_ref_3g}.

\begin{figure*}
\centering  
\includegraphics[width=18.cm]{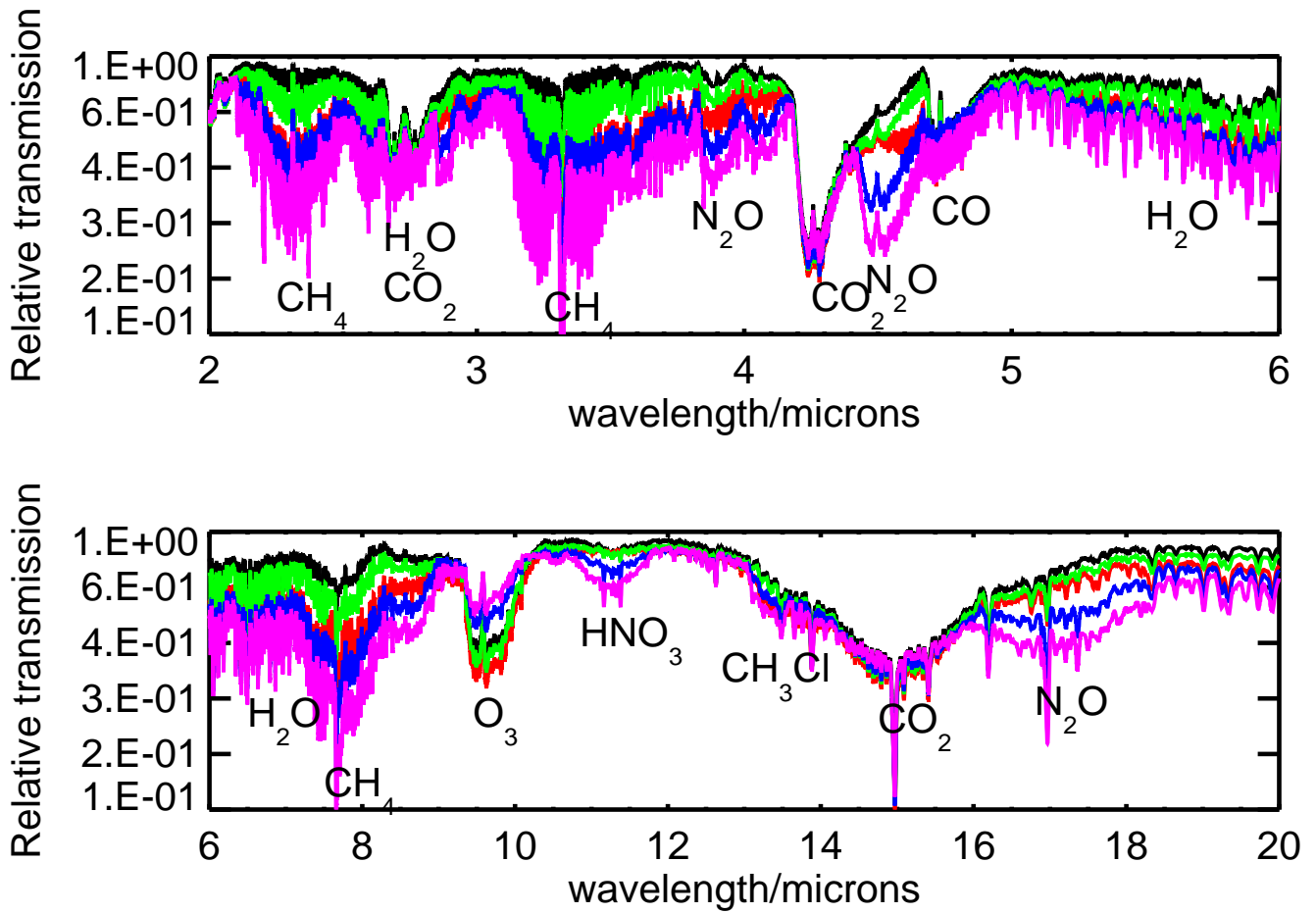}
\caption{Comparison of atmospheric transmission of a 1g Earth-like
  planet around the Sun (black), AD Leo (red), M0 (green), M5 (blue),
  and M7 (magenta) dwarf stars. }\label{reltrans_ref_1m}
\end{figure*}

The effect of changing the central star for the 1g scenarios is shown in Fig.
\ref{reltrans_ref_1m} for various M-dwarf type central stars. The
relative transmission spectrum (from Eq.~\ref{effheight} ) of the Earth is shown
in black.  Absorption from water, ozone, carbon dioxide, and methane
are clearly visible. Furthermore, the nitrous oxide 4.5 $\mu$m band and
17 $\mu$m band as well as a HNO$_3$ band at 11.5 $\mu$m can be
distinguished. Some interesting differences in the spectra are seen
for M dwarfs. The 15 $\mu$m band of CO$_2$ changes only by a small
amount from the Sun up to the M7 central star.  However, the
nitrous oxide, methane and water bands  continuously increase from
earlier (hotter) to later (cooler) M dwarfs. We note that N$_2$O
absorption can be as strong as the main CO$_2$ bands for the simulated
planet around the M7 star. In this case, the nitrous oxide would be a
strong biomarker signal.  The HNO$_3$ absorption basically
  disappears for AD Leo and is weak for the central M0 star case, but
 strongly  increases up to the quiet M7 star.

Ozone is strongest in AD Leo and increases somewhat from M0 to
  M3 for the quiet central stars. The ozone signal then, however,
  decreases again for planets around cooler central stars due to the
  reduced overall O$_3$ column. Thus, although atmospheric absorption
  signals in transmission may be easier to detect for planets orbiting
  cooler stars for most species, this may not be the case for ozone.


\begin{table*}
\centering
\caption{SNR for transmission spectra of an Earth-like planet at 10 pc\tablefootmark{a}\label{summarytransmearth}}
\begin{tabular}{c c c c c c c}\hline\hline
Stellar type & Molecule & $\lambda$ /$\mu$m & $R$=5 & $R$=20 & $R$=500 &
$R$=2000\\\hline
Sun & CH$_4$& 3.3& 7.2$\cdot 10^{-1}$& 5.2$\cdot 10^{-1}$& 1.1$\cdot 10^{-1}$& 5.4$\cdot 10^{-2}$\\
    & CO$_2$& 4.3& 1.15              & 1.10              & 2.4$\cdot 10^{-1}$& 1.2$\cdot 10^{-1}$\\
    & CH$_4$& 7.7& 4.9$\cdot 10^{-1}$& 3.3$\cdot 10^{-1}$& 7.1$\cdot 10^{-2}$& 3.5$\cdot 10^{-2}$\\
    & O$_3$ & 9.6& 5.2$\cdot 10^{-1}$& 4.6$\cdot 10^{-1}$& 9.2$\cdot 10^{-2}$& 4.5$\cdot 10^{-2}$\\
    & CO$_2$&15.0& 5.7$\cdot 10^{-1}$& 3.4$\cdot 10^{-1}$& 7.8$\cdot 10^{-2}$& 3.6$\cdot 10^{-2}$\\\hline
AD Leo & CH$_4$&3.3& 1.41            & 8.9$\cdot 10^{-1}$& 1.6$\cdot 10^{-1}$& 7.8$\cdot 10^{-2}$\\
    & CO$_2$& 4.3& 1.24              & 9.1$\cdot 10^{-1}$& 2.0$\cdot 10^{-1}$& 9.9$\cdot 10^{-2}$\\
    & CH$_4$& 7.7& 8.5$\cdot 10^{-1}$& 4.9$\cdot 10^{-1}$& 1.1$\cdot 10^{-1}$& 5.3$\cdot 10^{-2}$\\
    & O$_3$ & 9.6& 5.5$\cdot 10^{-1}$& 4.3$\cdot 10^{-1}$& 8.5$\cdot 10^{-2}$& 3.3$\cdot 10^{-2}$\\
    & CO$_2$&15.0& 5.4$\cdot 10^{-1}$& 3.1$\cdot 10^{-1}$& 6.8$\cdot 10^{-2}$& 3.3$\cdot 10^{-2}$\\\hline
M0  & CH$_4$& 3.3& 8.3$\cdot 10^{-1}$& 5.4$\cdot 10^{-1}$& 1.1$\cdot 10^{-1}$& 4.9$\cdot 10^{-2}$\\
    & CO$_2$& 4.3& 9.6$\cdot 10^{-1}$& 8.1$\cdot 10^{-1}$& 1.8$\cdot 10^{-1}$& 8.4$\cdot 10^{-2}$\\
    & CH$_4$& 7.7& 5.2$\cdot 10^{-1}$& 3.3$\cdot 10^{-1}$& 7.1$\cdot 10^{-2}$& 3.5$\cdot 10^{-2}$\\
    & O$_3$ & 9.6& 4.2$\cdot 10^{-1}$& 3.6$\cdot 10^{-1}$& 7.1$\cdot 10^{-2}$& 3.7$\cdot 10^{-2}$\\
    & CO$_2$&15.0& 4.5$\cdot 10^{-1}$& 2.8$\cdot 10^{-1}$& 5.9$\cdot 10^{-2}$& 2.9$\cdot 10^{-2}$\\\hline

M5& CH$_4$& 3.3  & 1.94              & 1.19              & 2.1$\cdot 10^{-1}$& 1.0$\cdot 10^{-1}$\\
    & CO$_2$& 4.3& 1.86              & 1.20              & 2.6$\cdot 10^{-1}$& 1.3$\cdot 10^{-1}$\\
    & CH$_4$& 7.7& 1.27              & 7.4$\cdot 10^{-1}$& 1.6$\cdot 10^{-1}$& 8.0$\cdot 10^{-2}$\\
    & O$_3$ & 9.6& 6.2$\cdot 10^{-1}$& 4.4$\cdot 10^{-1}$& 8.4$\cdot 10^{-2}$& 4.4$\cdot 10^{-2}$\\
    & CO$_2$&15.0& 7.2$\cdot 10^{-1}$& 4.2$\cdot 10^{-1}$& 9.2$\cdot 10^{-2}$& 4.5$\cdot 10^{-2}$\\\hline
M7& CH$_4$   &3.3& 2.66              & 1.59              & 2.9$\cdot 10^{-1}$& 1.4$\cdot 10^{-1}$\\
    & CO$_2$& 4.3& 2.47              & 1.40              & 3.0$\cdot 10^{-1}$& 1.5$\cdot 10^{-1}$\\
    & CH$_4$& 7.7& 1.78              & 1.02              & 2.2$\cdot 10^{-1}$& 1.1$\cdot 10^{-1}$\\
    & O$_3$ & 9.6& 7.7$\cdot 10^{-1}$& 4.7$\cdot 10^{-1}$& 9.1$\cdot 10^{-2}$& 4.7$\cdot 10^{-2}$\\
    & CO$_2$&15.0& 9.0$\cdot 10^{-1}$& 5.0$\cdot 10^{-1}$& 1.1$\cdot 10^{-1}$& 5.3$\cdot 10^{-2}$\\\hline

\end{tabular}
\hfill\\
\tablefoottext{a}{Integration time is a single transit duration.
Telescope configuration: JWST.}\\
\end{table*}

\subsubsection{Signal-to-noise-ratios}

\begin{figure}[!h]
\centering
  \includegraphics[scale=0.5]{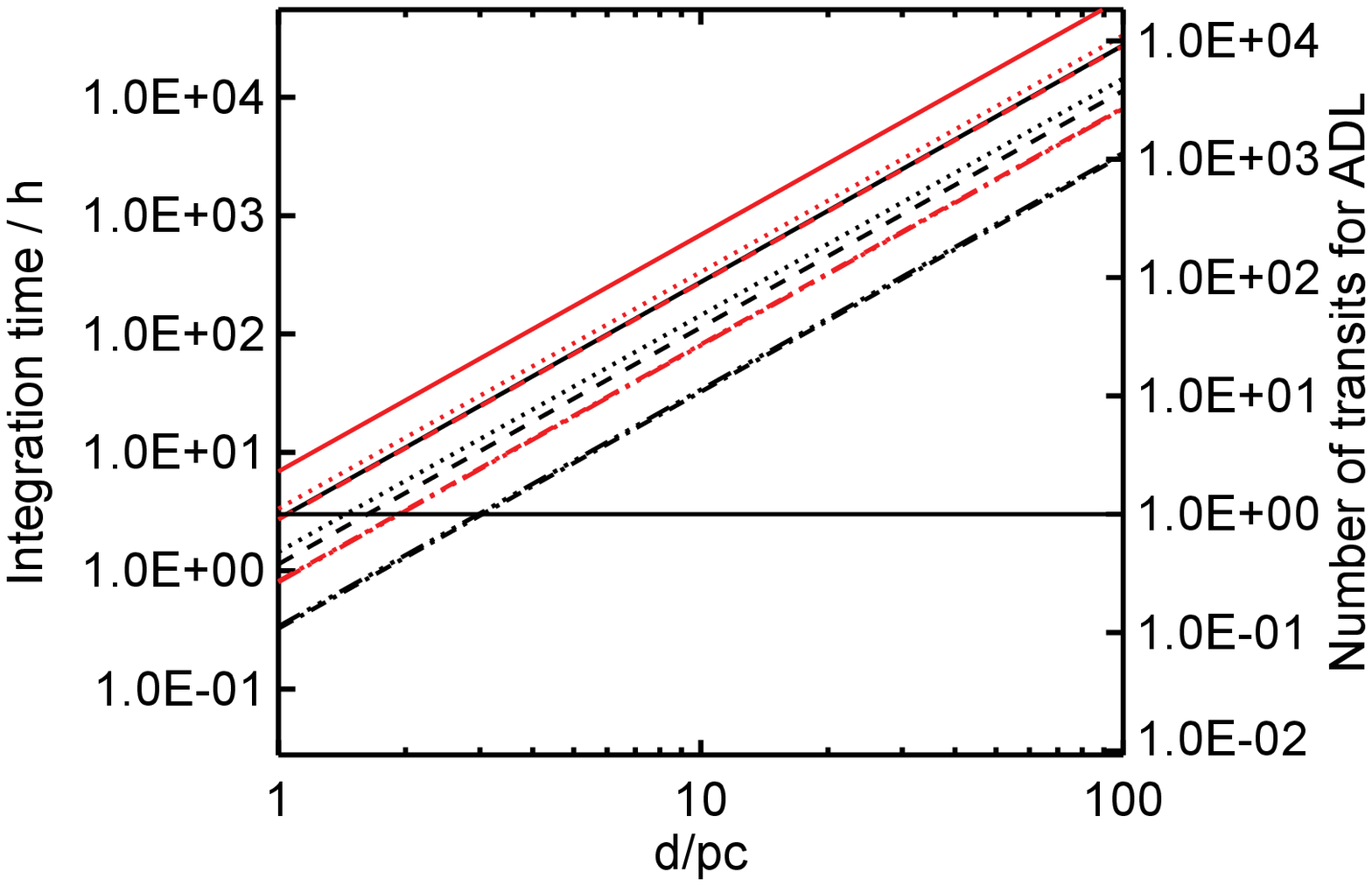}
  \includegraphics[scale=0.5]{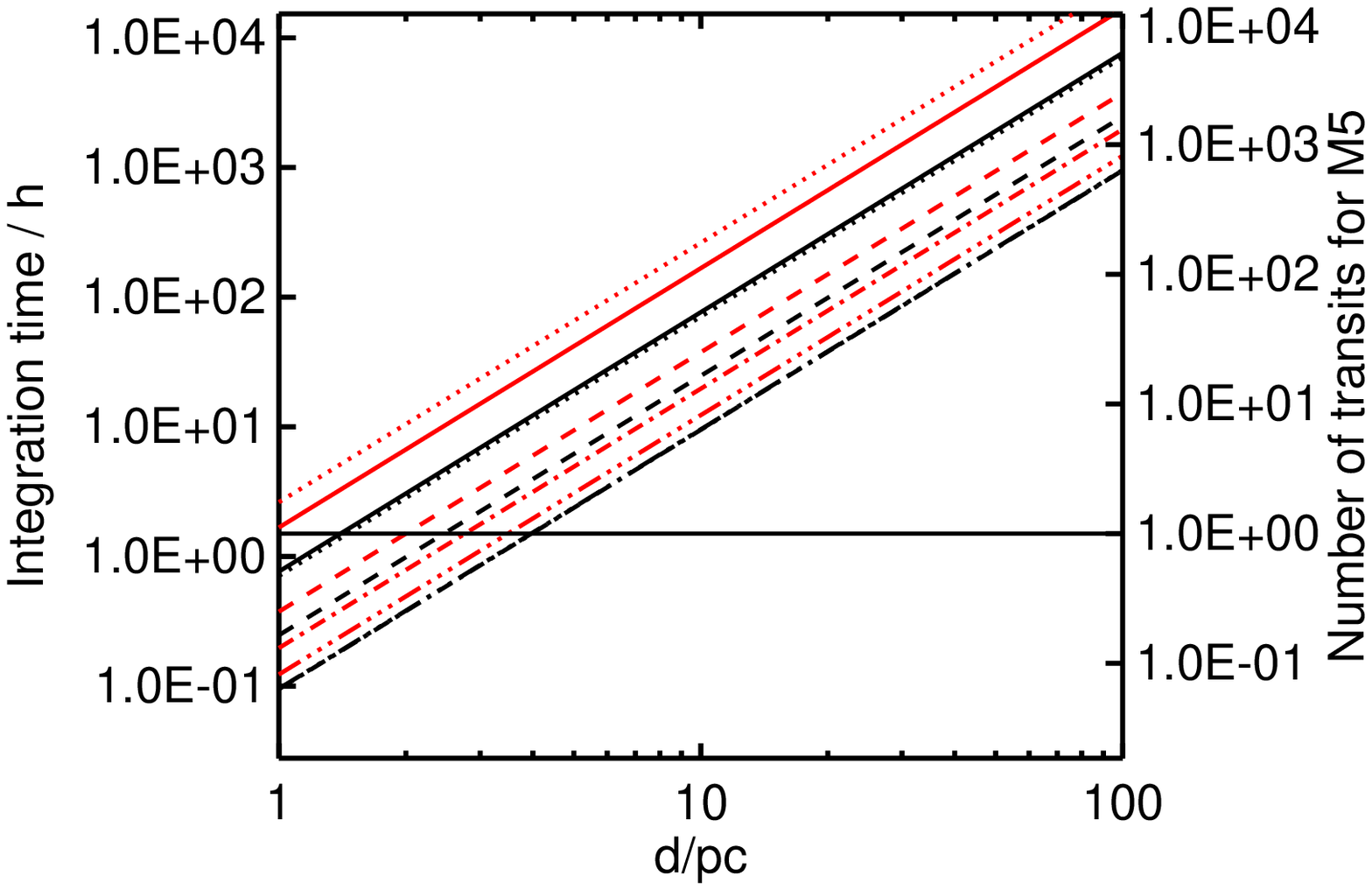}
  \caption{Transmission: Integration time needed for SNR=3 as a
    function of stellar distance of the 3.3 $\mu$m methane
    (dash-dot-dot-dot), 4.3 $\mu$m carbon dioxide (dash-dot), 7.7
    $\mu$m methane (dashed), 9.6 $\mu$m (dotted) ozone and the 15
    $\mu$m (solid) carbon dioxide spectral feature at $R=20$ for an
    Earth (black) and a 3g super-Earth (red) around AD Leo (top) and
    M5 (bottom). Telescope configuration is JWST.}
\label{transtimecomp}
\end{figure}

Table \ref{summarytransmearth} summarises the achievable SNR for
different absorption bands for transmission spectroscopy during a
single transit for Earth-sized planets around the Sun and M dwarfs. In contrast
to emission spectroscopy, transmission spectroscopy does not probe the
temperature structure of an atmosphere, but the chemical
composition. Accordingly high resolution is not explicitly
necessary. Unfortunately, at mid-IR wavelengths (e.g. 15 $\mu$m), the
stellar SNR is relatively low, which makes detection  harder than at
shorter wavelengths (e.g., ozone at 9.6 $\mu$m or methane at 3.3
$\mu$m). For near-IR absorption bands of CH$_4$ and CO$_2$ a SNR>1 can
be achieved in a single transit, thus providing the possibility to achieve
reasonable SNR if several transits can be co-added. Figure
\ref{transtimecomp} illustrates this concept. It shows integration
times as a function of required distance to achieve a SNR of 3.

Figure \ref{transtimecomp} implies that for close stars (closer than 5
pc) a JWST aperture-like telescope would be capable of performing
low-resolution transmission spectroscopy for Earth-like planets on all
bands for the simulated M5 dwarfs.  Earlier types of M dwarfs,
however, provide more challenging detection conditions.
To cross-check our results and identify the differences of the
approach used here, we compared our approach to that used by
\citet{Kaltenegger2009}. The difference is that we calculated the
atmospheric concentration profiles consistent with different M dwarf spectra, whereas \citet{Kaltenegger2009} used a fixed
modern-Earth reference profile. Both approaches yield quite similar
results for the CO$_2$ 15 $\mu$m band.  However, for e.g. the methane
and ozone absorption we find significant differences. This shows that
a consistent approach is necessary to take into account the effect of
stellar spectral energy distribution on atmospheric T-p profiles and
biomarker mixing ratios.


\section{Summary and conclusions}

We investigated the effect of the stellar energy flux
distribution of M-dwarf stars on atmospheric profiles of Earth-like
and super-Earth planets. In addition, IR emission and transmission spectra and SNRs were computed to investigate how this affects the spectral appearance and detectability of absorption bands.  We studied planets
with 1g and 3g (1-2 Earth radii) because this approximately corresponds
to the range from Earth to super-Earth planets. The planets were
placed at a distance from their host star where the stellar energy
input equals the energy of the Earth received by the Sun.  To
illustrate the detection capabilities for planets with biospheres, we 
arbitrarily assumed planets with modern Earth atmospheric
composition as an initial starting value. However, the actual
atmospheric composition was then re-calculated according to the
incident stellar spectral flux to obtain a consistent model of the
climatic conditions and composition of the atmospheres for these
planets.

Our results suggest that the different spectral energy distribution of M
dwarfs has a potentially large effect on the atmospheric T-p-profiles
and on the chemical profiles of H$_2$O, CH$_4$, O$_3$, N$_2$O, and
HNO$_3$. In particular the temperature inversion in the
mid-atmosphere disappears for planets around M dwarfs
(Fig. \ref{t-p-profile-star}). This was already found by
  \citet{Segura2005} for planets around active M-dwarf stars. For
  quiet stars with low UV flux we find, however, that an inversion can
  build up again in the planet's atmosphere owing to enhanced concentrations of water and methane
  in the mid-atmosphere.  In contrast to previous approaches, we did
  not fix the surface temperature to 288K through stellar flux adjustments
  in our simulations. Surface
  temperatures can be up to 20 K higher for planets around M dwarfs
  for the same total stellar energy input, but the simulated planets always
  retain their habitable conditions. The effect on atmospheric
structure and composition appears to be weaker when we varied gravity
from 1-3g compared with varying the stellar input spectra
(Fig. \ref{t-p-profile-g}).

We found systematic increases in the atmospheric abundances for
  water, methane, and nitrous oxide owing to the reduced photolysis
  rates associated with M-dwarf stars compared with the Sun. Abundances
  generally increase from hot to cool M dwarf stars.  For active M
  dwarfs, like AD Leo, photolysis heating rates are higher, and thus
  their related chemical processes are stronger than for the quiet M
  dwarfs of a similar type. Interestingly, planetary ozone abundances
  decrease for very cool, quiet central stars. This is caused by low
  photo\-lysis rates which reduce the Chapman ozone production. 
  In \citet{Grenfell2010} (paper II) we present a detailed
  analysis of the chemical responses of these atmospheres.

The main goal of this work is to present simulated spectra and
  SNR estimates for the modelled planetary systems. We therefore
  discussed simulated spectra for secondary and primary transit
  geometries. During secondary eclipse, the measured emission fluxes
  depend not only on the composition of the atmosphere, but also on
  its temperature structure. The influence of the decreasing stellar effective
  temperature of M0-M7 stars on the planetary atmospheres can be seen in our simulated planet
  spectra. As already pointed out by \citet{Segura2005}, absorption of
  O$_3$, CH$_4$, and N$_2$O is easier to detect for planets that orbit
  active M dwarfs. We confirm this result for AD Leo and also for
  quiet M-dwarf central stars up to about type M3.  For cooler central
  stars, however, the developing temperature inversion in the
  mid-atmosphere leads to reduced absorption signals, which almost
  level out for the coolest star considered (M7). Thus, although the
  abundances for most molecules continuously increase  from M0 to M7
  central stars, absorption features of atmospheric molecules are very
  difficult to detect in emission spectroscopy of planets that orbit
  very cool M dwarfs, which lead to potentially false negative detections
  of biomarkers in these cases.

Our simulations of transmission spectra during primary
  eclipses show that the relative transmission in spectral absorption
  bands of CO$_2$, H$_2$O, CH$_4$ increases from M0 to M7 dwarfs,
  hence increases towards cooler M central stars. We point out that
  the increasing relative transmission of the biomarker N$_2$O becomes
  comparable with the major CO$_2$ band for Earth-sized planets around M7 stars. Because N$_2$O is
  mainly produced from denitrifying bacteria on Earth, it is considered to be a good
  biomarker molecule. Unfortunately, for most planet scenarios the N$_2$O
  absorption bands are far too weak to be detected compared with
  the dominant atmospheric constituents. This may be different for
  planets around central stars with very low UV fluxes.

The situation is different for transmission spectra of the
  biomarker ozone. For active and hot M dwarfs ozone absorption also
  increases. However, for planets around cool M stars (M4-M7) the
  ozone absorption is reduced again because of its smaller atmospheric
  abundances caused by the low UV flux in quiet M dwarf stars.

Our simulations confirm that low spectral resolution (around R=20) is
sufficient to identify most spectral features in the spectrum of a
planet with an Earth-like atmosphere for IR emission as well as
transmission spectra. A characterisation of an atmospheric temperature
structure using IR emission spectra, e.g. by studying the presence of
a temperature inversion, would require much higher spectral
resolutions (of about >200).

Generally, emission spectra are studied in the IR wavelength
range. Our SNR calculations show that the detection of the broad
CO$_2$ absorption feature at 15 $\mu$m and the ozone band at 9.6
$\mu$m could be feasible for planets around very nearby M dwarfs with
exposure times of several hours for our modelled planet scenarios.

Transmission spectra obtained during primary eclipse of the planet, on
the other hand, favour shorter wavelength ranges owing to the increasing
stellar flux and therefore better SNR. In this case, detection of
e.g. methane at 3.3 $\mu$m may be possible when adding transits for
the nearest M dwarfs. Summing many transits with a stable instrument
could bring the measurements into a feasible range for future
observations.

Of course, the real atmospheric composition and total pressures of
super-Earth planets remain unknown until the first spectroscopic
detections can be made. Atmospheres even of planets similar to Earth
may differ in atmospheric mixing ratios, or show e.g. more extended
atmospheres. Clouds will additionally affect the detection of
absorption signals (see e.g. \citealp{Kitzmann2010b}). This will
result in different SNR from those computed here, where only particular
scenarios were studied. Investigations of a wide parameter range
would therefore be needed. Looking for yet undetected planets, it is
important to be open minded and take into account the unexpected,
e.g. when designing instruments. Consistent
modelling of climate and chemistry in the atmospheres of extra-solar
planets is important, even when possible only in exemplary cases,  for
understanding effects on T-p-profiles and on the composition and therefore
spectral appearance. An example is  the
  effect of the T-p-profiles for emission spectra of cool M dwarf
  stars, which in the extreme case leads to extremely shallow absorption signals and consequently false negative detections. Another example is the
  ozone signal in transmission spectra, where the ozone column is
  significantly reduced  for planets around very cool stars owing to
  chemical effects. For the main CO$_2$ 15 $\mu$m band, however,  the
differences are small when only considering detection of the broad
absorption band. Nevertheless, if future telescopes will be able to
observe terrestrial exoplanets at higher spectral resolution (R>100),
a consistent model approach for CO$_2$ is also mandatory to interpret
the temperature inversion peak of this band.


\begin{acknowledgements}
This research was partly supported by the Helmholtz
Gemeinschaft (HGF) through the HGF research alliance "Planetary
Evolution and Life".
We thank the anonymous referee for helpful comments.
A. Belu, F. Selsis and P. Hedelt acknowledge support from the European Research Council 
(Starting Grant 209622: E3ARTHs)
\end{acknowledgements}

\small{
\bibliographystyle{aa}
\bibliography{Bibo_1D}
}


\end{document}